\documentclass[%
 aip,
 jmp,%
 amsmath,amssymb,
 floatfix,
 reprint,%
]{revtex4-1}

\usepackage{geometry,amsmath,graphicx,hyperref}
\usepackage[english]{babel}
\usepackage{dcolumn}% Align table columns on decimal point
\usepackage{bm}% bold math

\newcommand{\tmop}[1]{\ensuremath{\operatorname{#1}}}
\newcommand{\tmtextbf}[1]{\text{{\bfseries{#1}}}}
\newcommand{\tmtextit}[1]{\text{{\itshape{#1}}}}
\newcommand{\nonconverted}[1]{\mbox{}}
\begin{document}

\title{Fast equilibrium reconstruction by deep learning on EAST tokamak}

\author{Jingjing Lu}
\affiliation{University of Science and Technology of China, Hefei, 230026, China}
\affiliation{Institute of Plasma Physics, Chinese Academy of Sciences, Hefei 230031, China}

\author{Youjun Hu}
\email{yjhu@ipp.cas.cn}
\affiliation{Institute of Plasma Physics, Chinese Academy of Sciences, Hefei 230031, China}

\author{Nong Xiang}
\email{nxiang@ipp.cas.cn}
\affiliation{Institute of Plasma Physics, Chinese Academy of Sciences, Hefei 230031, China}

\author{Youwen Sun}
\affiliation{Institute of Plasma Physics, Chinese Academy of Sciences, Hefei 230031, China}

\begin{abstract}
  A deep neural network is developed and trained on magnetic measurements
  (input) and EFIT poloidal magnetic flux (output) on the EAST tokamak. In
  optimizing the network architecture, we use automatic optimization in
  searching for the best hyperparameters, which helps the model generalize
  better. We compare the inner magnetic surfaces and last-closed-flux surfaces
  (LCFSs) with those from EFIT. We also calculated the normalized internal
  inductance, which is completely determined by the poloidal magnetic flux and
  can further reflect the accuracy of the prediction. The time evolution of
  the internal inductance in full discharges is compared with that provided by
  EFIT. All of the comparisons show good agreement, demonstrating the accuracy
  of the machine learning model, which has the high spatial resolution as the
  off-line EFIT while still meets the time constraint of real-time control.
\end{abstract}

{\maketitle}

\

\section{Introduction}

Reconstructing magnetic configuration using magnetic measurements is a routine
task of tokamak operation. There are many equilibrium solvers, e.g.,
EFIT{\cite{Lao_1985,Lao_1990,Lao_2005,EAST2009,JET1992,KSTAR2011}}, that can
do this kind of reconstruction by solving the Grad-Shafranov equation under
the constraint of magnetic measurements. In recent years, accumulation of data
resulting from these reconstruction practices, along with the development of
machine learning algorithms, software frameworks and computing power, have
made it possible to train deep neural networks to provide reconstructions as
accurate as those by EFIT. This has been demonstrated on
KSTAR{\cite{KSTAR_2020}} and DIII-D{\cite{DIII-D_2022}}.

On the EAST tokamak{\cite{Wan_2017}}, EFIT has been routinely used in tokamak
operations for more than ten years and substantial equilibrium data have been
accumulated{\cite{Qian2009,Li_2013,Luo2010,Wan_2023}}. In this paper, we
report the results of magnetic reconstruction by a deep neural network trained
on the magnetic measurements and EFIT reconstructed 2D magnetic poloidal flux.

There are two versions of EFIT used on EAST, one is for real-time control and
one for off-line analysis. The former is often restricted to lower accuracy
due to the time constraint of real time control, while the latter is of higher
accuracy. In this work, we use the off-line EFIT data in training the neural
network. The model trained this way have the higher accuracy as the off-line
EFIT while still meets the time constraint of the real-time control.

There are many hyperparameters in a neural network that usually need to be set
manually, such as number of hidden layers, units per layer, mini-batch size,
learning rate, number of epochs of training. In recent years, there appear
optimization libraries that can automatically set the values of these
hyperparameters. In this work, we use the Optuna optimization
framework{\cite{optuna_2019}} in setting the hyperparameters. The
hyperparameters found this way turn out to be much better than our previously
manually set ones in terms of the model accuracy. The size of the network
architecture found by the automatic hyperparameter tuning turns out to be
relatively small (with less than 2 million parameters). This small size allows
for very fast equilibrium construction that can be easily deployed.

The input to the network is limited to only the magnetic measurements. The
output of the network is the 2D poloidal magnetic flux function, $\Psi (R, Z)
\equiv A_{\phi} R$, which is related to the poloidal magnetic field, $B_R$ and
$B_Z$, by
\begin{equation}
  \label{7-14-p1} B_R = - \frac{1}{R} \frac{\partial \Psi}{\partial Z},
\end{equation}
\begin{equation}
  \label{6-9-a4} B_Z = \frac{1}{R} \frac{\partial \Psi}{\partial R},
\end{equation}
where $(R, \phi, Z)$ are the cylindrical coordinates. The 2D contours of
$\Psi$ in $(R, Z)$ plane correspond to the magnetic surfaces. We compare the
inner magnetic surfaces and the LCFSs with those given by EFIT, in order to
evaluate the accuracy of $\Psi$ predicted by the network. We also calculate
the normalized internal inductance, $l_i$, which is a quantity that is solely
determined by $\Psi$ and thus can reflect how accuracy the predicted $\Psi$
is. The time evolution of the internal inductance in full discharges is
compared with that provided by EFIT. All of the comparisons show good
agreement, demonstrating the accuracy of the machine learning model, which has
the high spatial resolution as the off-line EFIT while still meets the time
constraint of real-time control.

The rest of this paper is organized as follows. Section
\ref{2_data_collection} presents how the data are collected and normalized.
Sec. \ref{3_NN} explains the structure of our neural network and how the
hyperparameters are chosen by automatic optimization. In section
\ref{4_results}, we test the predicting capability of the trained network.
Section \ref{23-5-18-p4} discusses a small network used to predict some
volume-averaged quantities, namely the plasma stored energy $W_{\tmop{mhd}}$,
normalized toroidal beta $\beta_N$, and edge safety factor $q_{95}$. A brief
summary is given in section \ref{5_summary}.

\section{Data collection and normalization}\label{2_data_collection}

\begin{figure}[h]
  \resizebox{0.45\columnwidth}{!}{\includegraphics{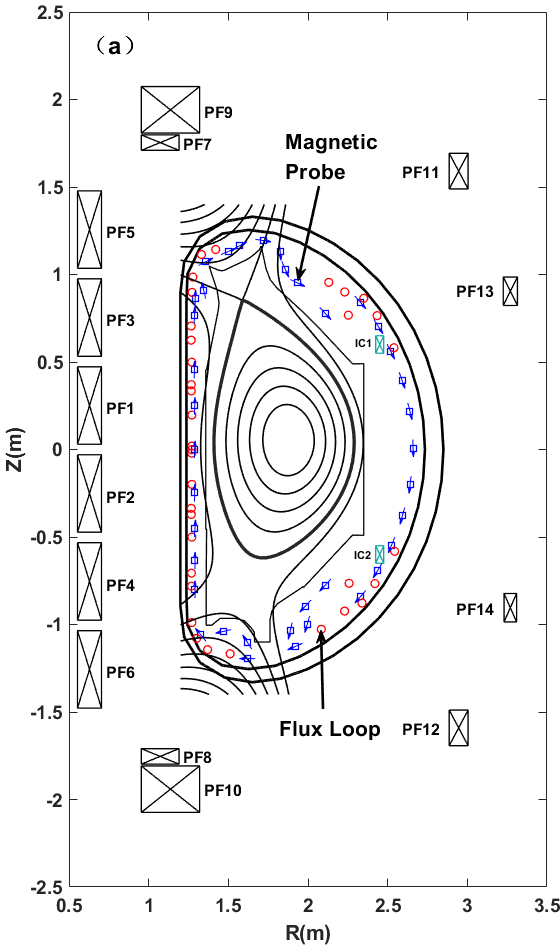}}\resizebox{0.48\columnwidth}{!}{\includegraphics{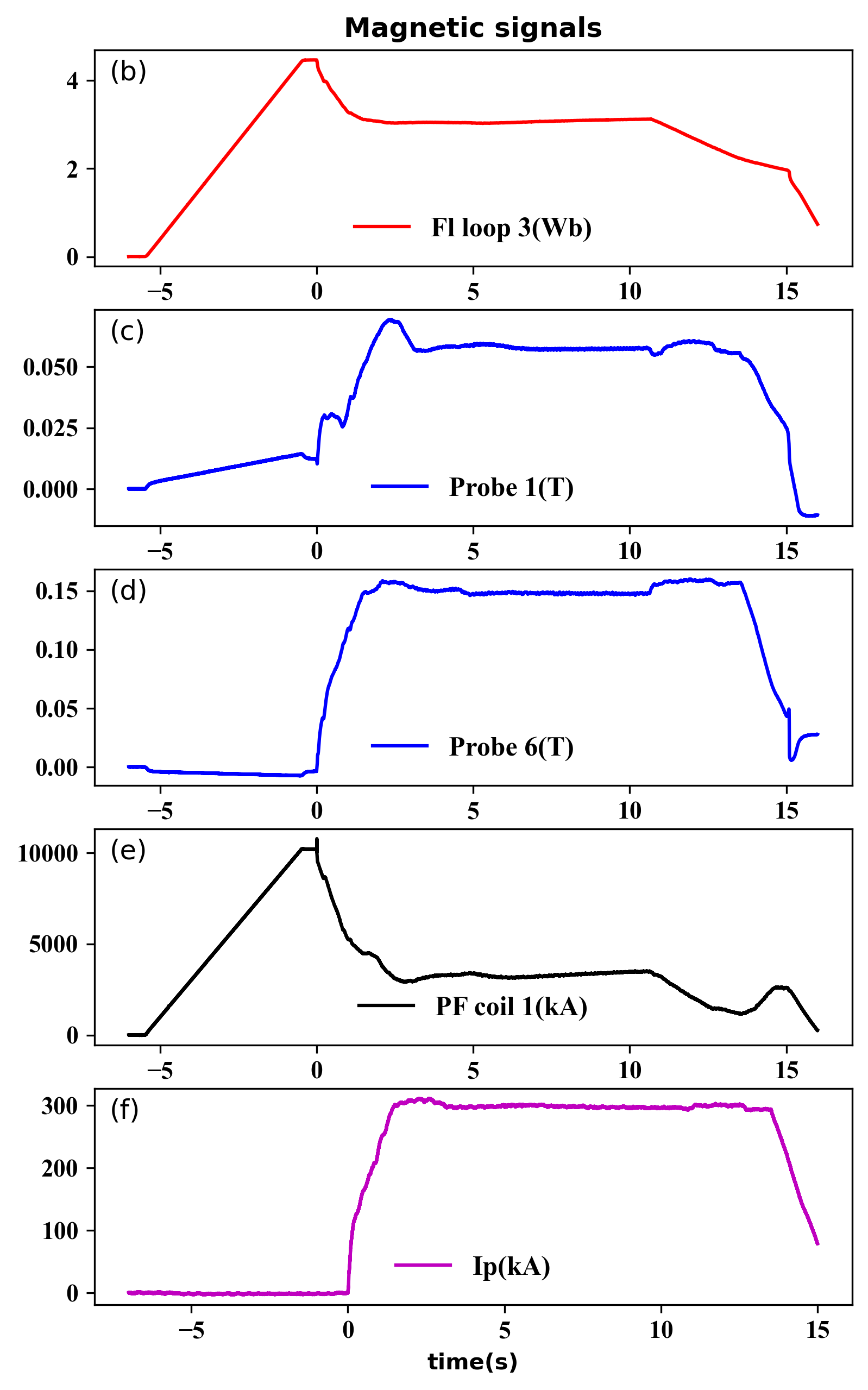}}
  \caption{\quad\label{figure2.1}Left: location of the magnetic probes, flux
  loops, and poloidal coils on EAST. A typical time evolution of some of the
  magnetic measurements from EAST discharge 113019 are plotted in the right
  panel. There are total 38 magnetic probes measuring the equilibrium poloidal
  magnetic field, and only 34 of them are working and are used in this work.}
\end{figure}

Figure \ref{figure2.1}a illustrates the poloidal locations of the magnetic
measurements used as inputs to our model. A typical time evolution of some of
the magnetic measurements from EAST discharge 113019 are plotted in Figure
\ref{figure2.1}b-f.

The inputs (features) to the neutral network (NN) are 84 magnetic
measurements: 35 poloidal magnetic flux ($\Psi_{\tmop{FL}}$) values measured
by flux loops, 34 equilibrium poloidal magnetic field (MP) values measured by
magnetic probes, 14 poloidal field (PF) coil currents and 1 plasma current
(I$_p$) measured by a Rogowski loop.

The outputs (targets) of the neutral network are the values of the poloidal
magnetic flux $\Psi$ at $R \times Z = 129 \times 129 = 16641$ spatial
locations. The $\Psi$ used in the training process is computed by off-line
EFIT and are downloaded from EAST MDSplus server (mds.ipp.ac.cn). The input
and the output signals are interpolated to the same time slices before they
are fed to the NN.

The inputs and outputs are summarized in Table \ref{table}.

\begin{table}[h]
  \resizebox{\columnwidth}{!}{
  \begin{tabular}{llll}
    \hline
    Signal & Measure method & Signal meaning & Num. of values\\
    \hline
    Input &  &  & 84\\
    \hline
    $\Psi_{\tmop{FL}}$ & Flux loop & Poloidal magnetic flux & 35\\
    MP & Magnetic probe & Poloidal magnetic field & 34\\
    PF & Rogowski loop & Poloidal field coil current & 14\\
    I$_p$ & \tmtextit{}Rogowski loop & Plasma current & 1\\
    \hline
    Output &  &  & 16641\\
    \hline
    $\Psi$(R,Z) & EFIT & Poloidal magnetic flux & 16641\\
    \hline
  \end{tabular}}
  \caption{ \ \label{table}The inputs and outputs of the model.}
\end{table}

The data used in training, validation and testing process were downloaded from
the EAST MDSplus server by using Python API, which scans a series of
discharges and automatically skips discharges where necessary signals are
missing. Specifically, we scan every 5 discharges among all the discharges
spanning from \#114000 to \#117000, resulting in total 45,544 equilibria (time
slices). These discharges are from experiments performed in one EAST campaign
from June to July in 2022. This range is casually chosen with no particular
criterion, except that we prefer recent discharges and avoid old discharges
because locations of some magnetic probes were changed in previous campaigns.
The auxiliary heating methods on EAST used in this campaign include neutral
beam injection (50-70keV Deuterium beam), lower-hybrid waves (2.45GHz and
4.6GHz), electron cyclotron waves (140GHz), and ion cyclotron waves
(25-70MHz). Typical values of total heating source power are between 4-10MW.

\begin{figure}[h]
  \resizebox{0.9\columnwidth}{!}{\includegraphics{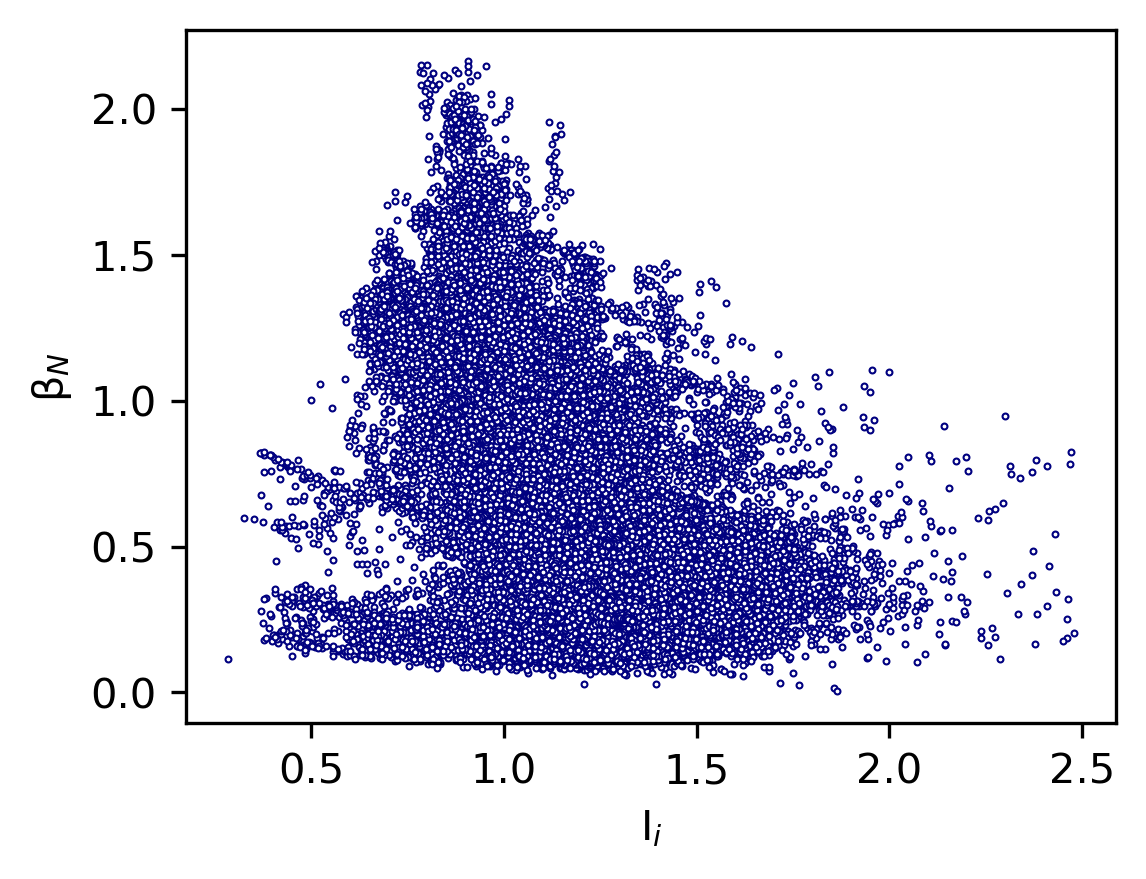}}
  \caption{ \label{li_}Distribution of the EFIT equilibrium data (training set
  + validation set+ testing set ) in $(l_i, \beta_N)$ plane.}
\end{figure}

Figure \ref{li_} is the distribution of the EFIT equilibrium data in $(l_i,
\beta_N)$ plane, where $l_i$ is the normalized internal induction and
$\beta_N$ is the normalized plasma beta.

The collected data are split into three sets: training set (81\%), validation
set (9\%), and testing set (10\%), where training set is used in training the
NN, validation set is used in monitoring potential overfitting and tuning
hyperparameters, and testing set is used in testing the predicting capability
of the trained model.

Figure \ref{figure2.1}b-f shows that there is a difference of up to six orders
of magnitude in the values of the input signals. In order to eliminate scale
differences among features, we use the min-max normalization method to
normalize the input data. The general formula is given by

\begin{equation}
  x' = \frac{x - x_{\min}}{x_{\max} - x_{\min}}
\end{equation}
where $x $ is the original value of the feature, $x'$ is the normalized value,
$x_{\min}$ and $x_{\max}$ are respectively the minimal and maximal value of a
feature in the data sets excluding the testing set. The $x_{\min}$ and
$x_{\max}$ obtained here are then used to normalize the input data in the
testing set when doing prediction using the trained NN.

\begin{figure}[h]
  \resizebox{0.9\columnwidth}{!}{\includegraphics{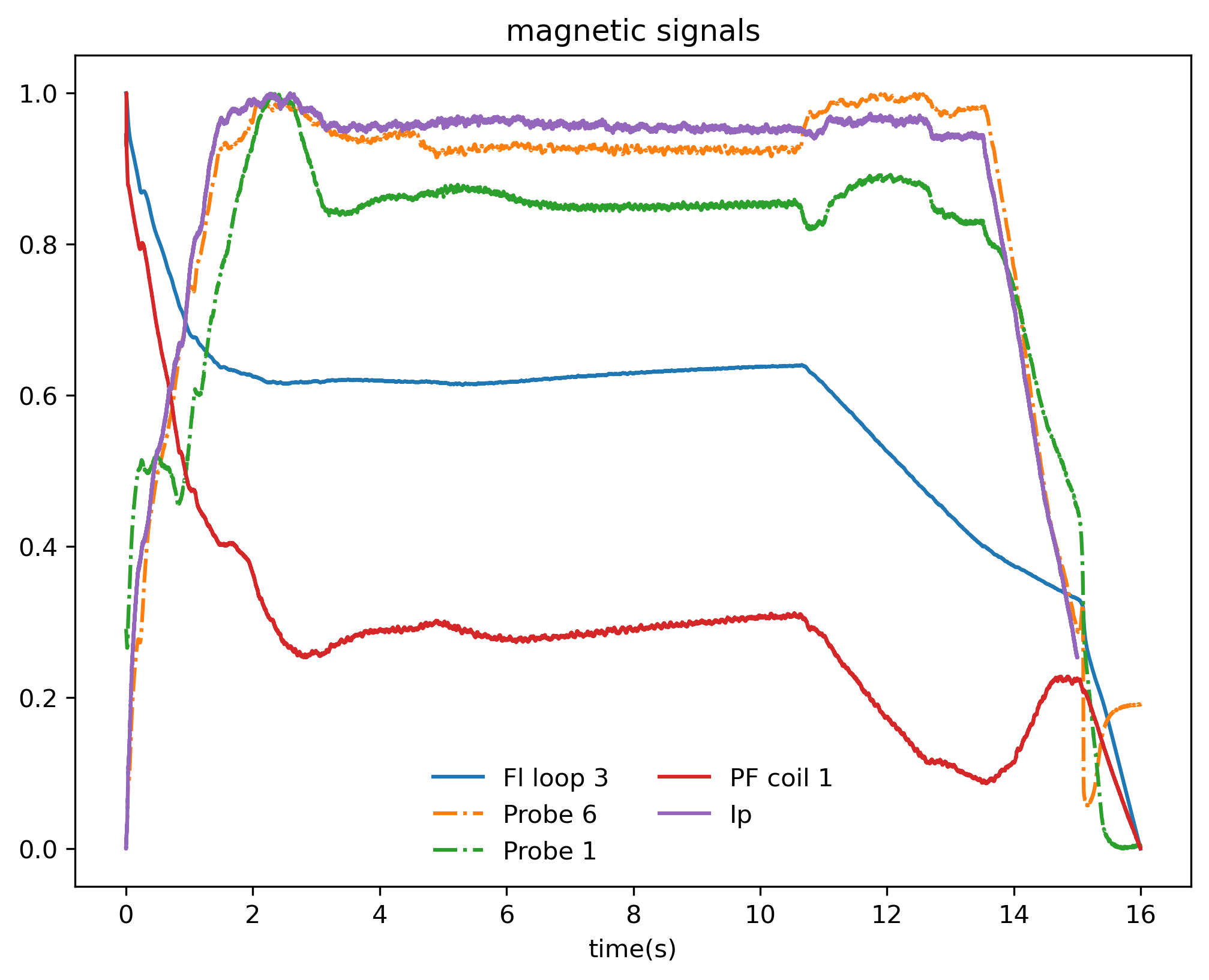}}
  \caption{\label{normalization}Normalized signals corresponding to those in
  Fig. \ref{figure2.1}b-f. The sudden change in the poloidal magnetic flux (Fl
  Loop 3) after 10s is brought about by the PF1 coil current, which is
  actively adjusted by the plasma control system to provide Ohm field to
  maintain a constant plasma current.}
\end{figure}

Figure \ref{normalization} plots the time evolution of the normalized input
signals corresponding to those in Fig. \ref{figure2.1}b-f.

The magnitude of the output ($\Psi$ in SI units) is near 1, so no
normalization is applied to it.

\section{Model architecture and automatic hyperparameter tuning}\label{3_NN}

An artificial neural network is a kind of computational network, which usually
consists of multiple layers: input layer, one or more inner layers (known as
hidden layers), and a layer of outputs. Each layer is made of units. Each unit
in the computing layers (hidden and output layers) receives information and
process the information using some linear transform (matrix multiplication)
and some nonlinear transform (activation function).

\begin{figure}[h]
  \resizebox{0.9\columnwidth}{!}{\includegraphics{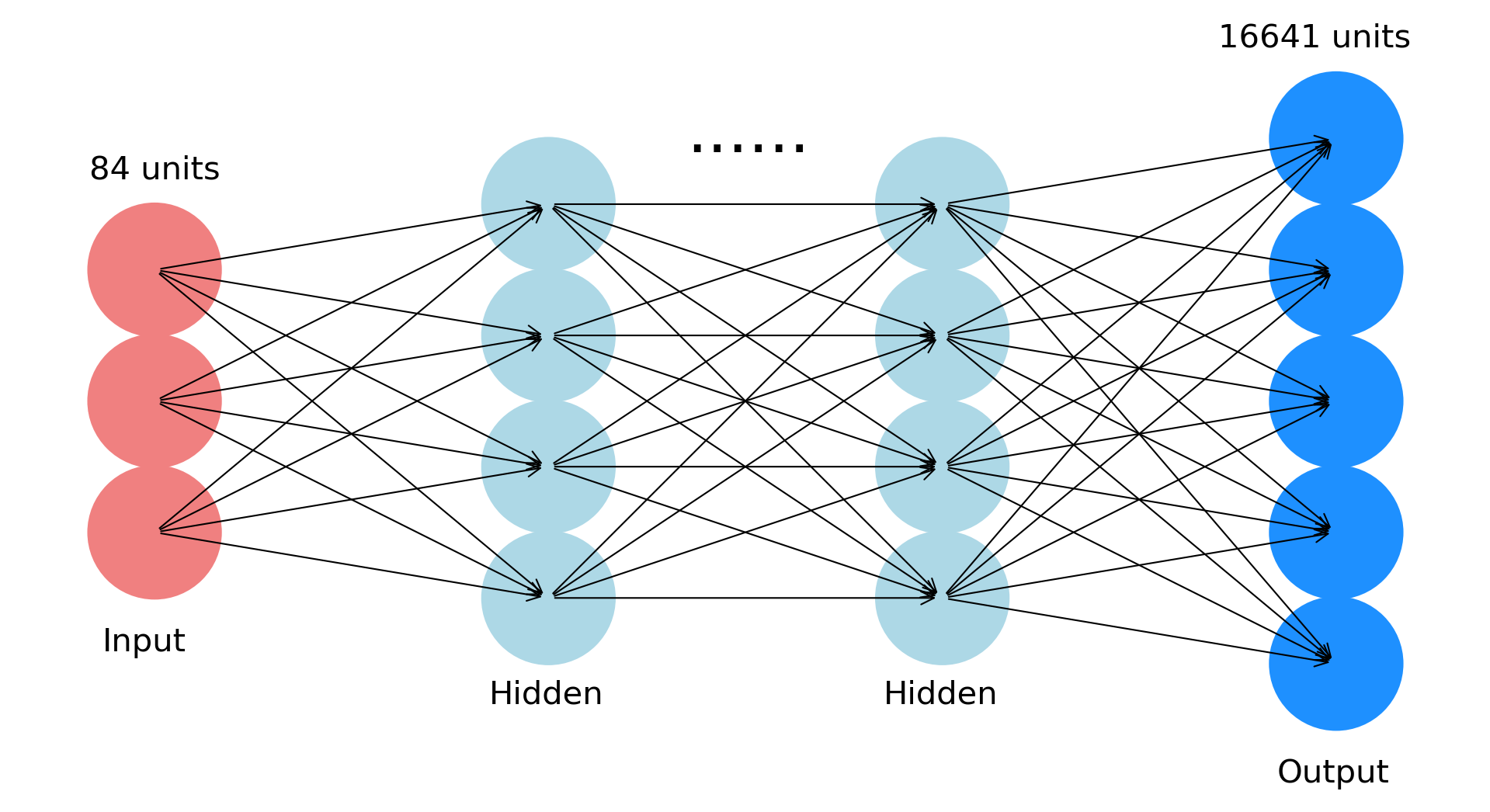}}
  \caption{ \label{NN}Fully connected feed-forward neural network used in this
  work.}
\end{figure}

A fully-connected feed-forward network showed in figure \ref{NN} is used here
to predict the poloidal magnetic flux $\Psi$ based on the magnetic
measurements. Here ``fully-connected'' means that each unit of a computing
layer receives information from all the units of the previous layer.
``Feed-forward'' means that information move in only one direction (from the
input layer to the hidden layers, and to the output layers), no cycles or
loops, and no intra-layer connections.

\

Each unit (neuron or node) in the computing layers has trainable parameters,
often called weights and biases. Denote the output of the $j^{\tmop{th}}$
neuron in $l^{\tmop{th}}$ layer by $a_j^l$, then a neural network model
assumes that $a_j^l$ is related to the $a^{l - 1}$ (output of the previous
layer) via
\begin{equation}
  \label{22-3-9-e2} a^l_j = \sigma \left( \sum_k w_{j k}^l a^{l - 1}_k + b^l_j
  \right),
\end{equation}
where $w^l_{j k}$ and $b_j^l$ are the weight and bias, the summation is over
all neurons in the $(l - 1)^{\tmop{th}}$ layer, and $\sigma$ is a function
called activation function. The weights and biases will be adjusted in the
training process by gradient descent methods to reduce the loss (cost or
error) function, which is defined in this work as
\begin{equation}
  \label{22-3-9-e1} L (\mathbf{w}, \mathbf{b}) \equiv \frac{1}{2 n} \sum_{i =
  1}^n \| \mathbf{y}_i - \hat{\mathbf{y}}_i \|^2,
\end{equation}
where $\mathbf{y}_i$ is the EFIT poloidal magnetic flux and
$\hat{\mathbf{y}}_i$ is the NN output, and the summation is over all the
samples in the training set. The loss function in Eq. (\ref{22-3-9-e1}) is the
mean squared error (MSE). The loss function measures the derivation of the
approximate solution away from the desired exact solution. So the goal of a
learning algorithm is to find weights and biases that minimize the loss
function. To minimize the loss function over $(\mathbf{w}, \mathbf{b})$ using
the gradient descent method, we need to compute the partial derivatives
$\partial L / \partial w^l_{j k}$ and $\partial L / \partial b^l_j$, which can
be efficiently computed by the well known back-propagating
method{\cite{Rumelhart1986,nielsen2015neural}}. The back-propagating algorithm
and the corresponding gradient descent method are the core algorithms in all
deep learning software frameworks.

Besides the trainable parameters, there are various hyperparameters in a NN
that usually need to be set manually, such as number of hidden layers, units
per layer, activation function, NN optimizers, learning rate, batch size,
number of epochs of training. In recent years, there appear automatic
optimization libraries that can search for the best combination of
hyperparameters. In this work, we use the Optuna optimization
framework{\cite{optuna_2019}} in setting the hyperparameters. Optuna automates
the hyperparameter optimization process by defining a search space of
hyperparameters and exploring the space using efficient searching algorithms.
The tree-structured Parzen estimator (TPE) algorithm is used in this work.
This algorithm models the relationship between hyperparameters and their
corresponding performance metrics and makes efficient decisions on which
hyperparameters to try next.

\begin{figure}[h]
  \resizebox{0.9\columnwidth}{!}{\includegraphics{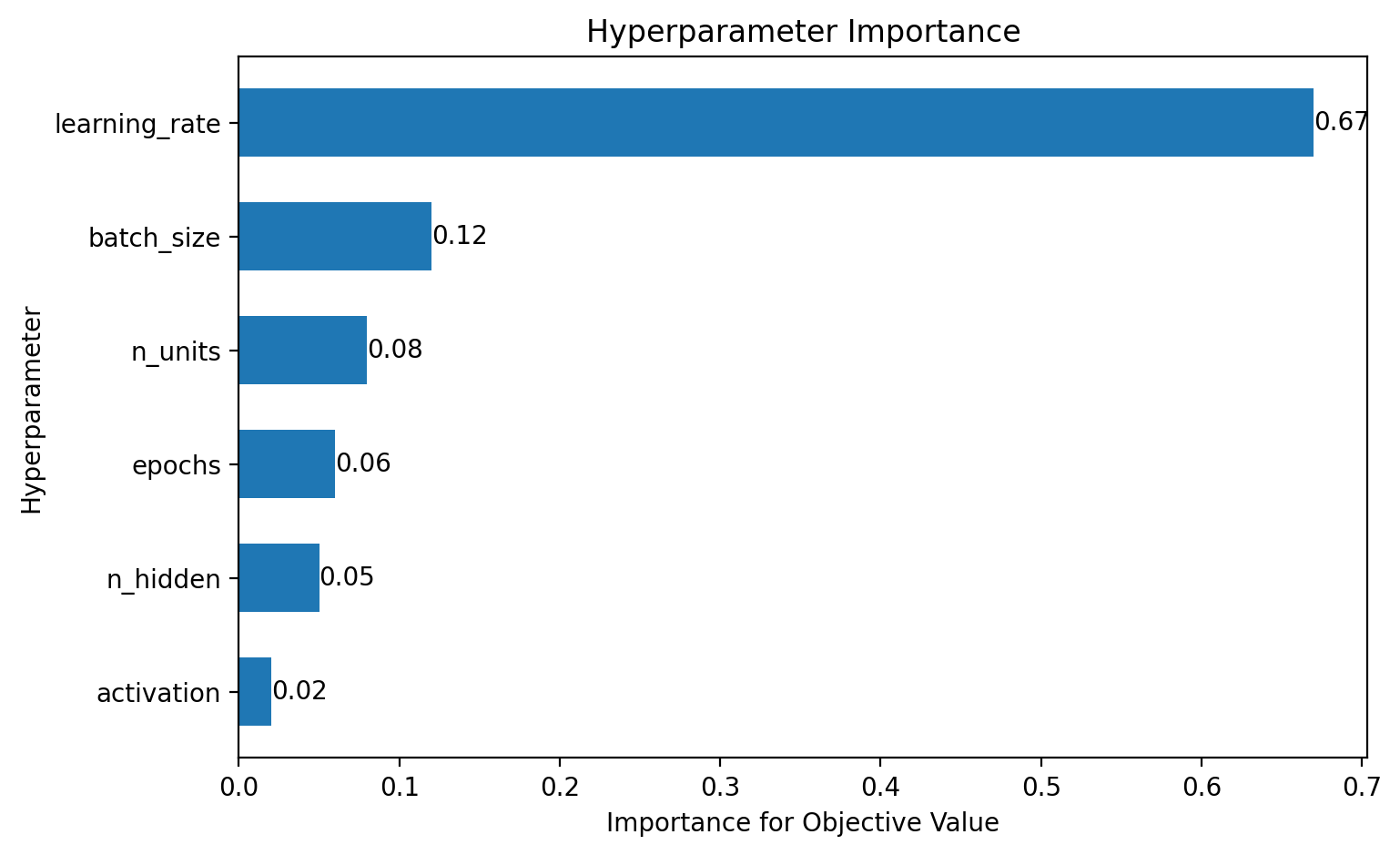}}
  \caption{ \label{23-3-24-p3}Hyperparameter importance distribution. Model
  accuracy is not sensitive to activation functions for this case.}
\end{figure}

\begin{figure}[h]
  \resizebox{0.9\columnwidth}{!}{\includegraphics{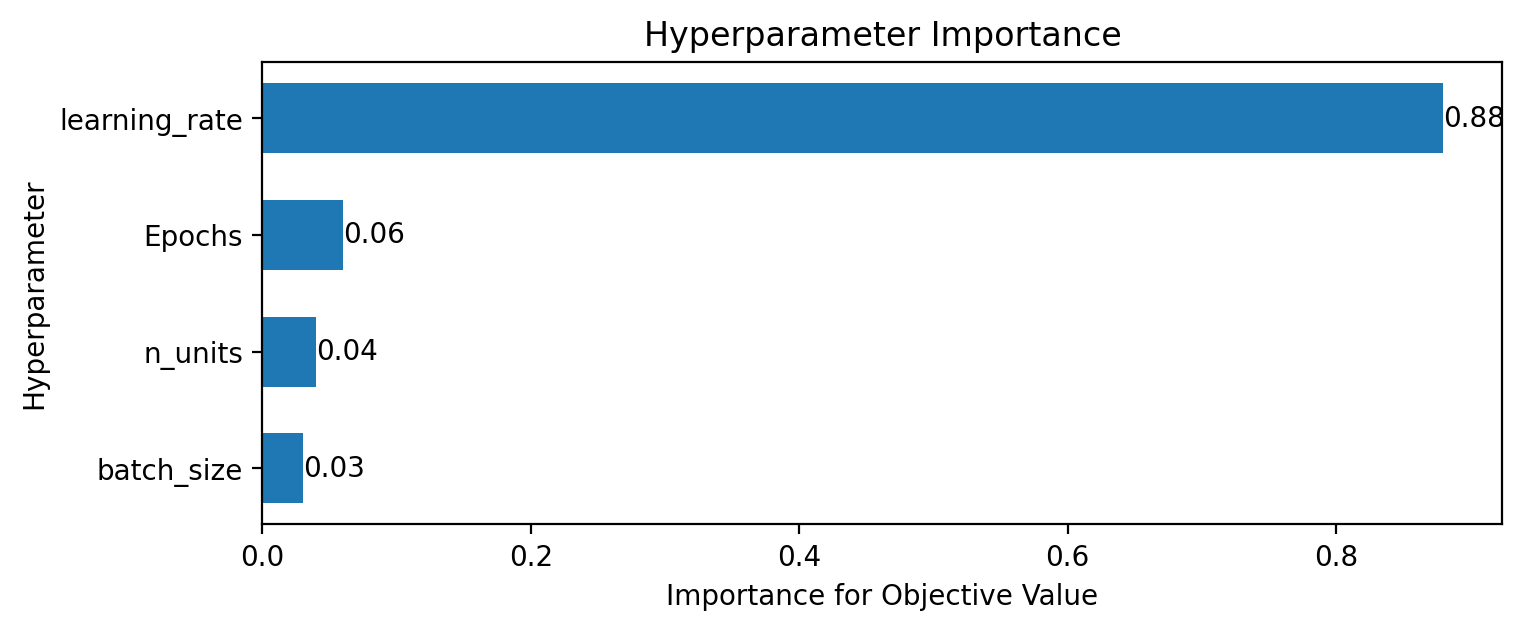}}
  \caption{\quad\label{23-3-24-p1}Relative importance of the hyperparameters
  in determining the model accuracy in the fine tuning.}
\end{figure}

Optuna automate the selection of the best hyperparameter combination. After
multiple experiments, we have found that the model accuracy is not sensitive
to the number of hidden layers, activation functions, and optimizers (an
example showing the relative importance of these hyperparameters is given in
Fig. \ref{23-3-24-p3}). Therefore, these hyperparameters are fixed in the fine
tuning step, in order to improve the speed of the model selection process, and
explore more hyperparameter regimes to which the model may be sensitive. For
other hyperparameters, we use Optuna framework to find the optimal combination
of hyperparameters. Relative importance of these hyperparameters are shown in
Fig. \ref{23-3-24-p1}. The above results indicate that learning rate is the
dominant factor that determines the model performance.

The final values of the hyperparameters used in the model are shown in Table
\ref{hyper_set}.

\begin{table}[h]
    \resizebox{\columnwidth}{!}{
  \begin{tabular}{c|c|c}
    \hline
    Hyperparameter & Meaning & Final values\\
    \hline
    n\_layers* & Number of hidden layers & 4\\
    n\_units & Number of nodes per hidden layer & 86\\
    Activation* & Activation function & tanh\\
    Optimizer* & Optimizer type & Adam\\
    $\eta$ & Learning rate & $2.26 \times 10^{- 5}$\\
    Loss* & Loss function & MSE\\
    batch\_size & Number of samples used in a step & 16\\
    Epochs & Number of epochs & 97\\
    \hline
  \end{tabular}}
  
  \
  
  * Fixed hyperparameters during fine tuning
  \caption{\quad\label{hyper_set}Final values of hyperparameters of the model.
  The hyperparameters with an asterisk (*) are fixed during the fine tuning.
  The above activation function refers to that used in the hidden layers. For
  the output layer, the linear activation function is used.}
\end{table}

\begin{figure}[h]
  \resizebox{0.9\columnwidth}{!}{\includegraphics{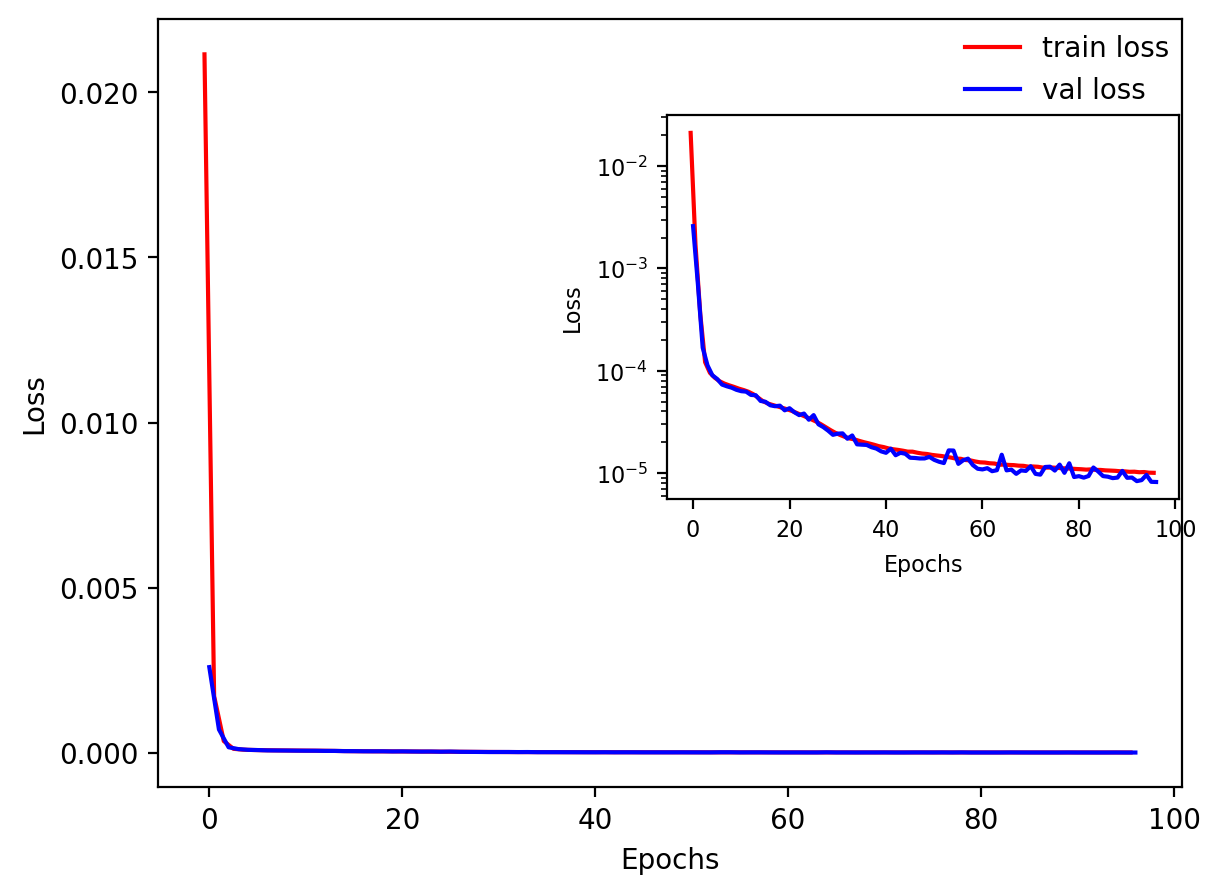}}
  \caption{ \label{23-3-16-p1}Training history showing the training and
  validation loss function values versus the training epochs. One epoch
  corresponds to go through all the samples in the training set. In the
  mini-batch stochastic gradient descent method used here, each gradient
  descent step uses randomly selected 16 samples (a mini batch), and the loss
  function values shown here are obtained by summation over a mini batch of
  samples in the training set or validation set.}
\end{figure}

The network are constructed and trained using Keras \& TensorFlow2
{\cite{chollet2015keras,t2015}}, which is a broadly adopted open source deep
learning framework in industry and research community. Figure.
\ref{23-3-16-p1} plots loss function values as a function of the training
epochs. The loss function is also evaluated on the validation set, which
serves as a monitor for the possible overfitting. The validation loss follows
the same trend as the training loss, indicating no overfitting.

\

\section{Performance of the neural network}\label{4_results}

\subsection{Performance of the model on testing set}

After the model is trained on the training set, we assess its prediction
capability on the data that are not seen by the training process. To evaluate
the reconstruction quality, we employ three widely adopted metrics: the
Pearson correlation coefficient $r$ (definition is given in appendix \ref{r}),
the coefficient of determination $R^2$ (definition is given in appendix
\ref{r2}), and the peak signal-to-noise ratio (PSNR) (definition is given in
appendix \ref{PSNR}).

Figure \ref{4.1}a plots the NN prediction of the poloidal flux $\Psi_{N N}$
vs. EFIT results $\Psi_{\tmop{EFIT}}$ for the testing set (total 4555
equilibria, each with 16641 values). The Pearson correlation coefficient $r$
and the coefficient of determination $R^2$ are also shown in the figure, which
are very close to 1, indicating a strong predictive capability. Figure
\ref{4.1}b plots the distribution of the correlation coefficient $r$ between
NN predictions and EFIT results for each equilibrium of the 4555 equilibria in
the testing set. The results indicate that the majority of the values are
greater than 0.998, indicating good correlation between NN prediction and EFIT
result for each equilibrium.

\begin{figure}[h]
  \resizebox{0.45\columnwidth}{!}{\includegraphics{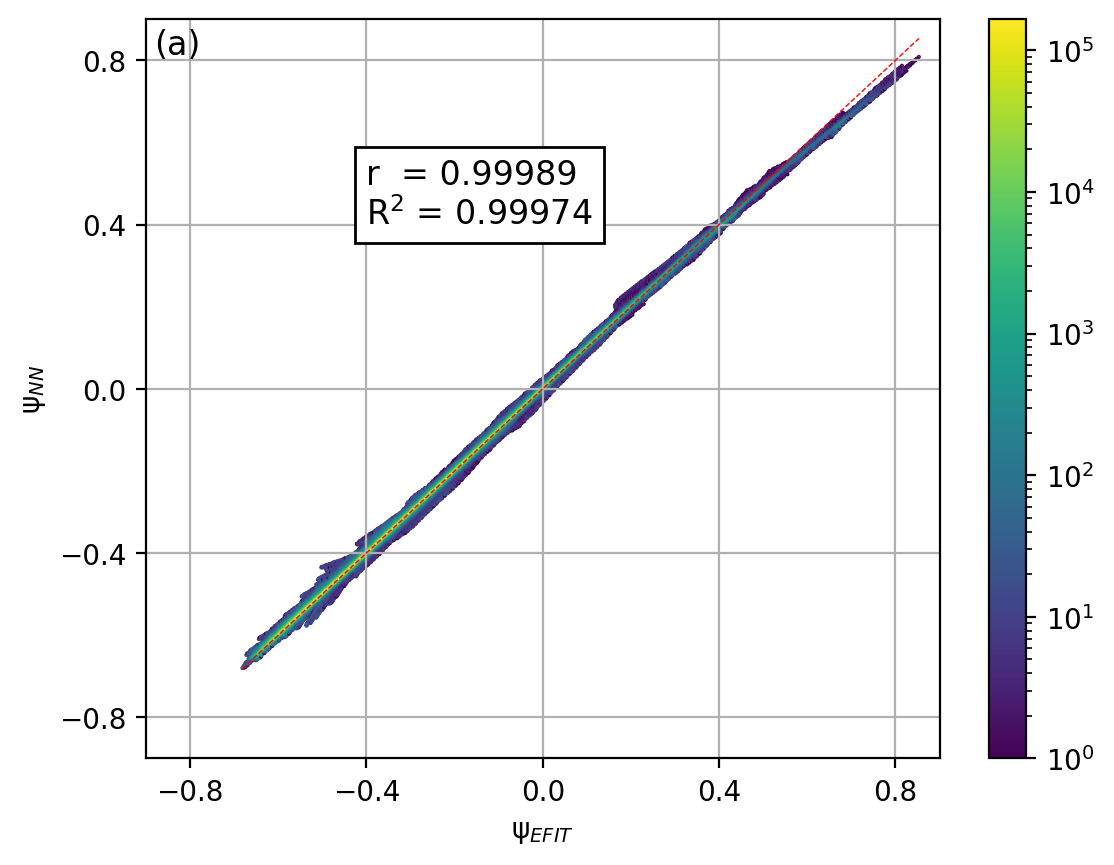}}
  \resizebox{0.45\columnwidth}{!}{\includegraphics{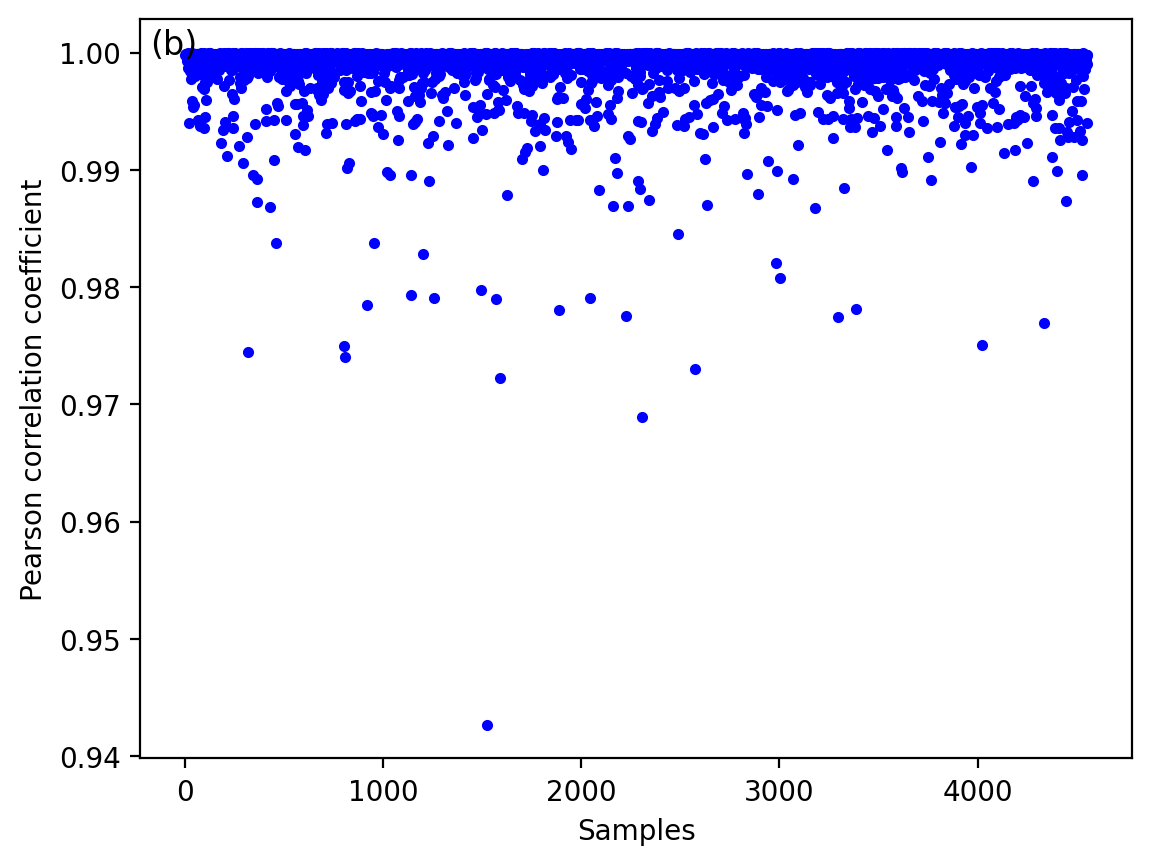}}
  \caption{ \label{4.1}(a) Neural network prediction of the poloidal flux
  $\Psi_{N N}$ vs. EFIT results $\Psi_{\tmop{EFIT}}$ for the testing set
  (total 4555 equilibria). Red dashed line is the $y = x$ line. The total
  number of points shown here is $4555 \times 129 \times 129$. Color
  represents density of data points. (b) The distribution of the correlation
  coefficient for each equilibrium of the 4555 equilibria in the testing set.}
\end{figure}

\begin{figure}[h]
  \resizebox{0.9\columnwidth}{!}{\includegraphics{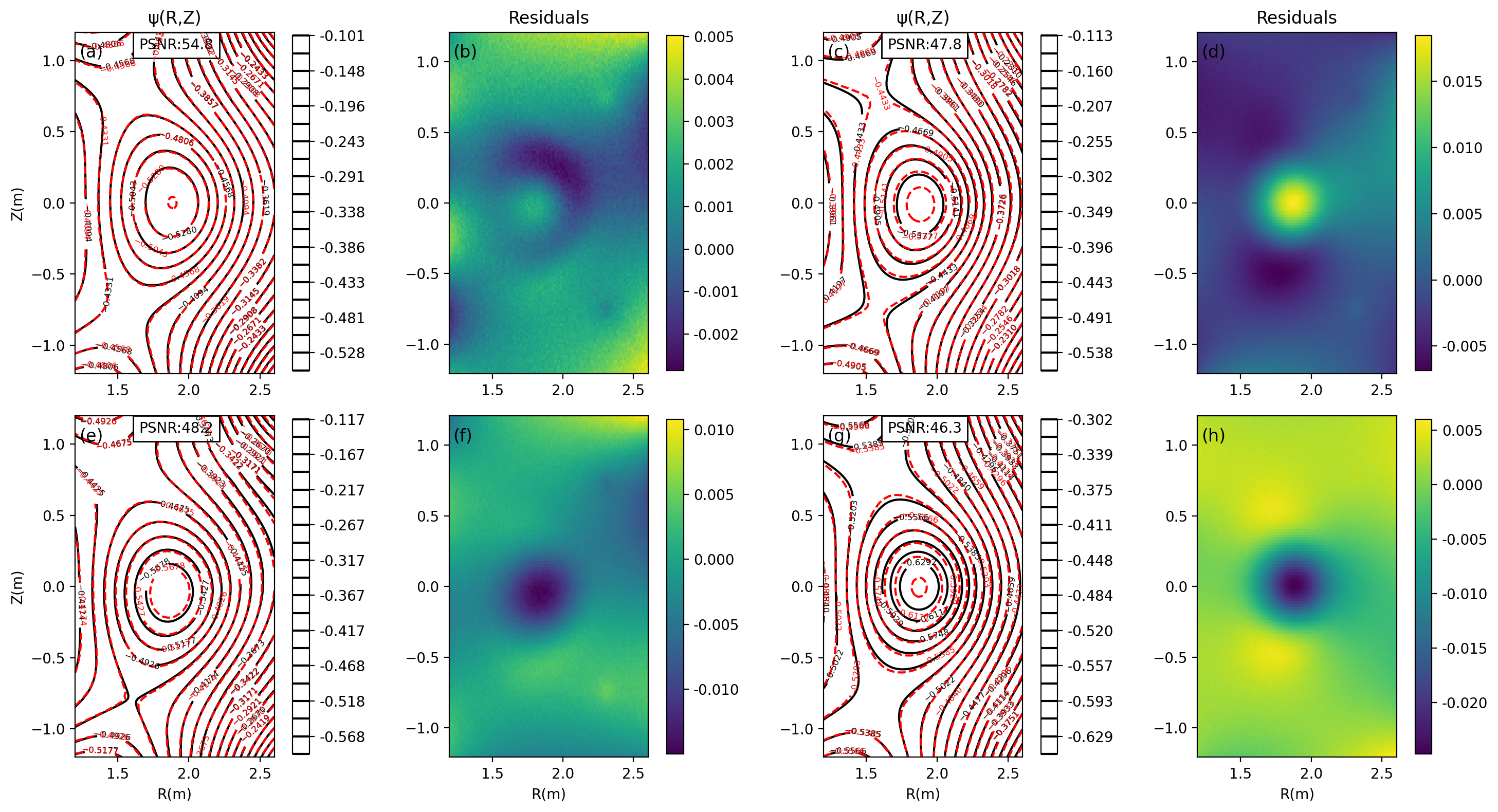}}
  \caption{ \label{4.2}Panels (a), (c), (e) and (g) compare the magnetic
  surfaces from EFIT (solid black lines) with those from the NN (dashed red
  lines) for four randomly selected time slices from the 4555 testing samples.
  Panels (b), (d,) (f) and (h) show the corresponding normalized residua
  [$\Psi$(R,Z)  \nonconverted{minus}
  $\Psi_{}^{\tmop{NN}}$(R,Z)]/max(\textbar$\Psi$ (R,Z)\textbar).}
\end{figure}

To test the accuracy of the model in predicting the plasma magnetic surface,
we compare the 2D contours of the poloidal magnetic flux predicted by the NN
with those given by EFIT. The results are shown in figures \ref{4.2}(a), (c),
(e) and (g), where the NN predictions of $\Psi$ contours are overlaid on the
$\Psi$ contours of EFIT. It displays four randomly selected samples from the
4555 equilibria in the testing set (the four displayed samples may not
necessarily come from the same discharge). Since our reconstruction results
take the form of images with resolution determined by the spatial grid points,
it is also useful to use $\tmop{PSNR}$ in evaluating the reconstruction
quality of the magnetic surface. The values of PSNR for the four equilibrium
are shown in the figure.

\

\

\begin{figure}[h]
  \resizebox{0.9\columnwidth}{!}{\includegraphics{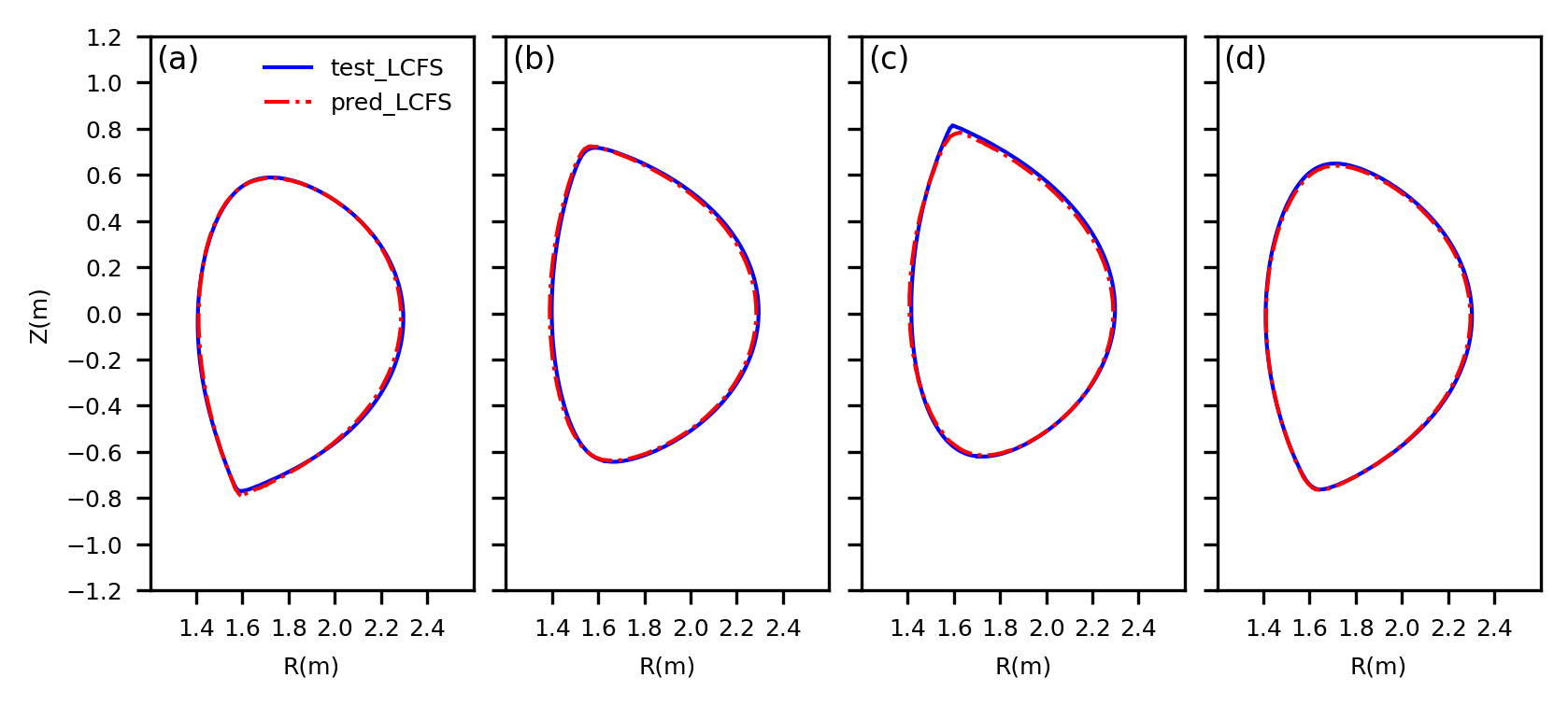}}
  \caption{ \label{23-3-21-a1}Comparison of LCFS between the NN prediction and
  EFIT result. The four panel correspond to the four equilibria shown in fig.
  \ref{4.2}. In locating the LCFS from the NN prediction, we analyze a series
  of $\Psi$ contours and determine the outermost contour that are near the
  magnetic separatrix.}
\end{figure}

To further assess the accuracy of the mode, we locate the LCFSs predicted by
the NN model and compare them with those given by EFIT. The LCFSs
corresponding to the four equilibrium of Fig. \ref{4.2} are shown in Fig.
\ref{23-3-21-a1}, which indicates that the NN and EFIT results are in good
agreement. Minor discrepancies appear near the X points.

\subsection{Performance of the model on four complete discharges}

In this section, we arbitrarily select 3 full discharges that are not in the
dataset used above to examine the time evolution of the magnetic configuration
during an entire discharge (from ramp-up to flat-top then to ramp-down).

Besides the plasma currents $I_p$, we also calculate the normalized internal
inductance $l_i$ (definition is given in Appendix \ref{23-3-21-a2}), which is
a quantity that is solely determined by $\Psi$ and thus can further reflect
how accuracy the predicted $\Psi$ is. We plot the time evolution of $l_i$ and
compare it with the EFIT results. By doing this, we can assess the accuracy of
the NN in predicting the time evolution of some key volume-integrated
quantities characterizing magnetic configuration.

\begin{figure}[h]
  \resizebox{1.0\columnwidth}{!}{\includegraphics{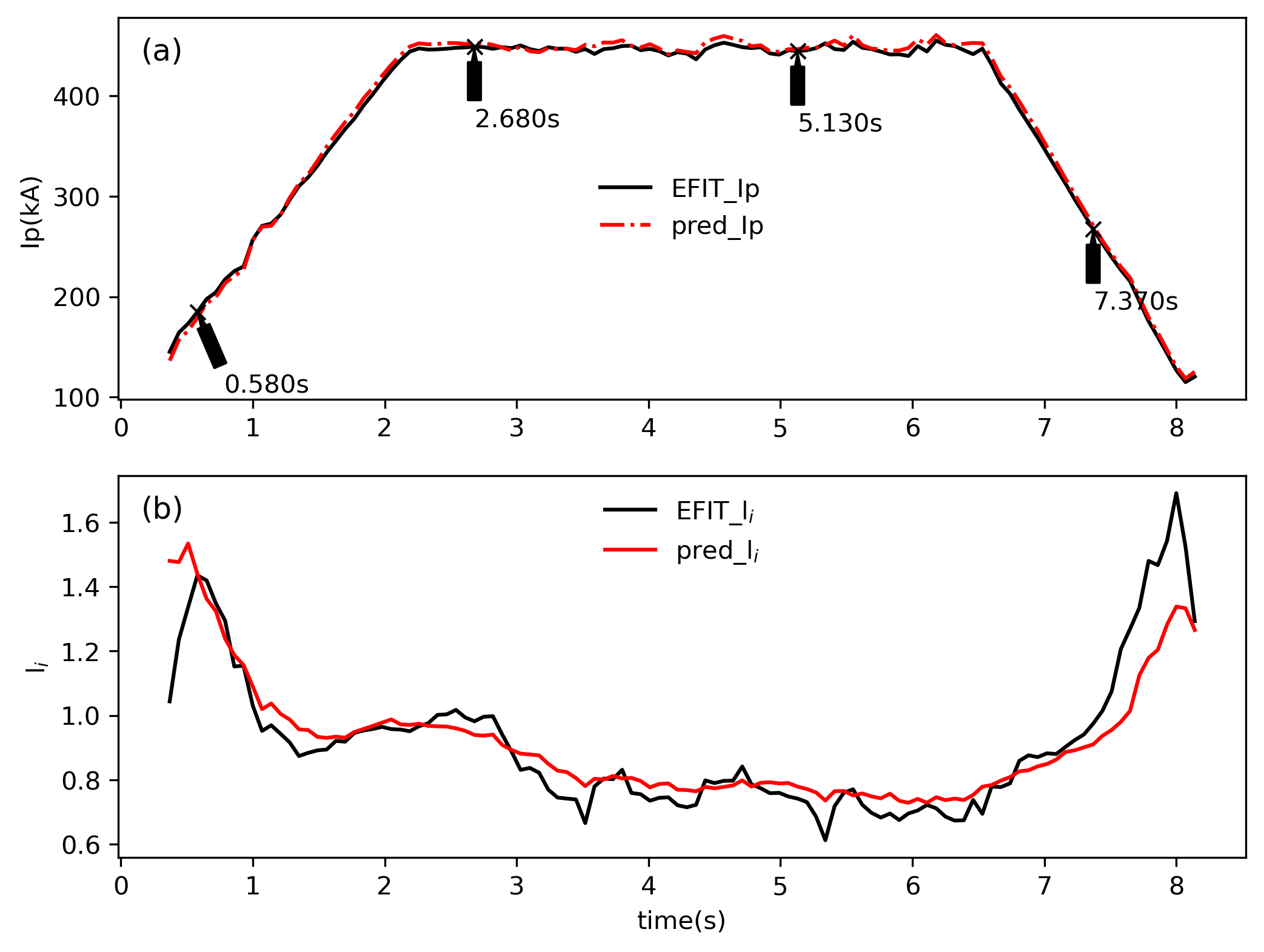}}
  \caption{ \label{4.4} (a) Comparison of time evolution of plasma current
  given by the NN and EFIT. (b) Comparison of time evolution of $l_i$ given by
  the NN and EFIT. Four time slices are indicated on the graph, which are time
  slices selected for the magnetic configuration comparison shown in Figs.
  \ref{4.5}-\ref{4.6}.}
\end{figure}

Figure \ref{4.4} compares the time evolution of $I_p$ and $l_i$ predicted by
the NN and that by EFIT for discharge \#113388.

\

\begin{figure}[h]
  \resizebox{1.0\columnwidth}{!}{\includegraphics{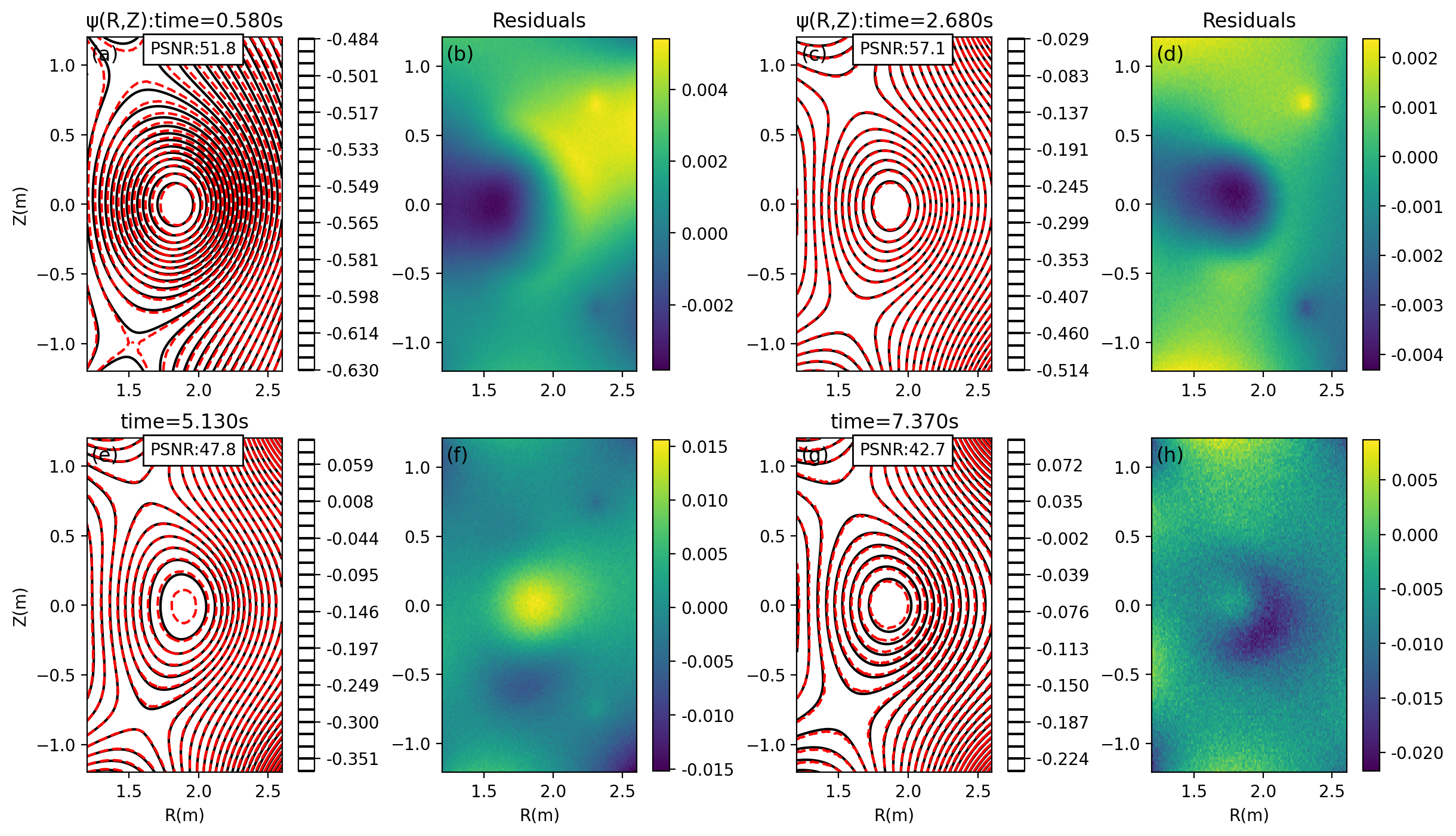}}
  \caption{ \label{4.5}Panels (a), (c), (e) and (g) compare the magnetic
  surfaces from EFIT (solid black lines) with those from the NN (dashed red
  lines) for discharge \#113388. $\Psi$ at four different time slices during
  the discharge, 0.580s (early ramp-up), 2.680s,, 5.130s (flat top), 7.370s
  (ramp-down) \ are shown. Panels (b), (d,) (f) and (h) show the corresponding
  relative error \ $(\Psi_{\tmop{EFIT}} - \Psi_{\tmop{NN}}) / \max (|
  \Psi_{\tmop{EFIT}} |)$.
  }
\end{figure}

Figure \ref{4.5} compares the contours of $\Psi$ given by the NN model and
that given by EFIT at 4 time slices (indicated in Fig. \ref{4.4}) in discharge
\#113388. The results indicate the relative error between the NN and EFIT
results is less than $2\%$.

\

\begin{figure}[h]
  \resizebox{1.0\columnwidth}{!}{\includegraphics{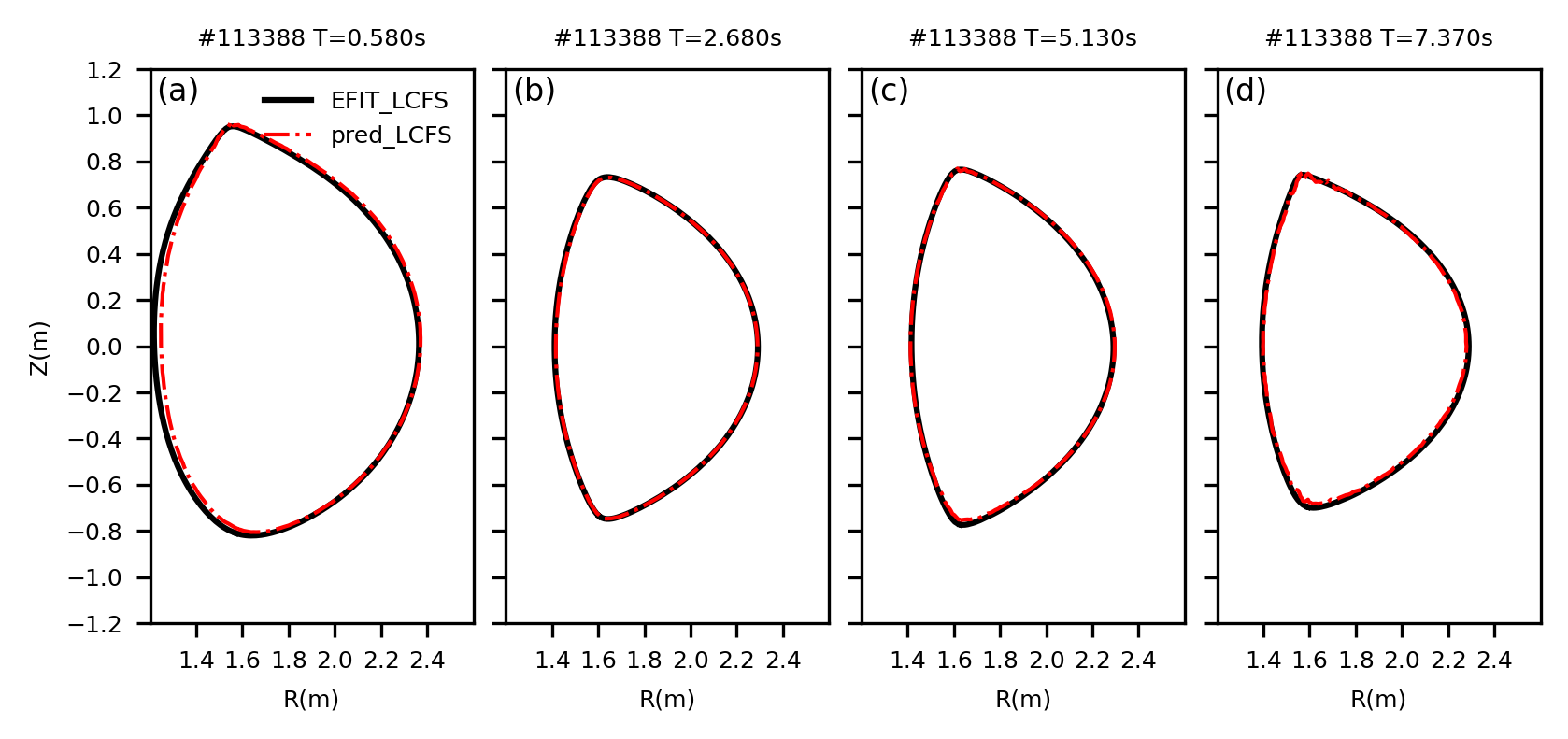}}
  \caption{ \label{4.6}NN reconstructions of the LCFS (dotted-dashed red) for
  EAST shot \#113388 overlaid against the EFIT LCFS (solid black). LCFSs from
  four different times in the discharge \#113388 are shown.}
\end{figure}

Figure \ref{4.6} compares the LCFSs given by the NN model and that given by
EFIT at 4 time slices in discharge \#113388. The results show good agreement
between the two models. Minor differences usually appear in the ramp up/down
phase, and near the X-points.

\

\

\begin{figure}[h]
  \resizebox{1.0\columnwidth}{!}{\includegraphics{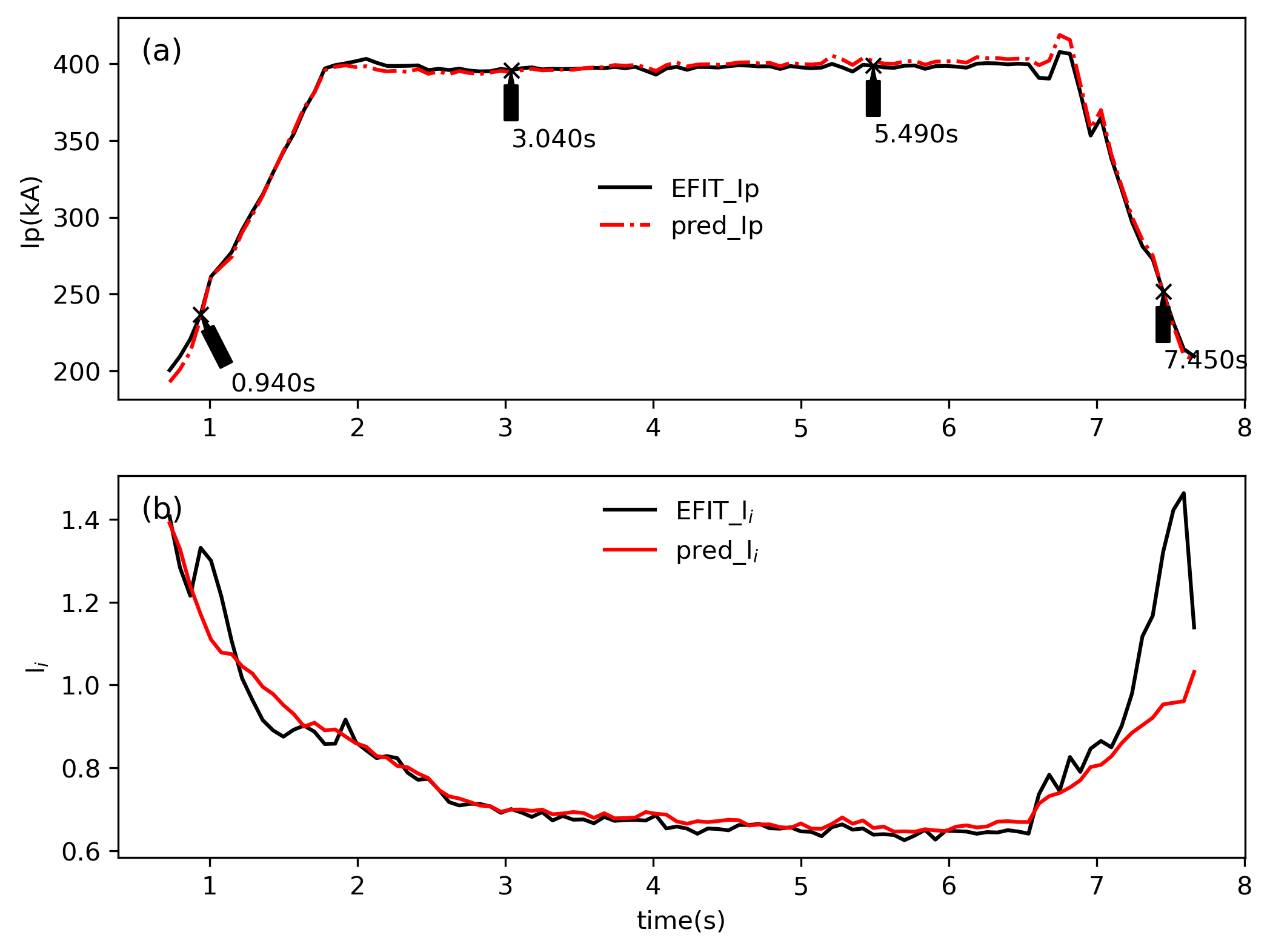}}
  \caption{ \label{4.7}Same as figure \ref{4.4}, except that the discharge is
  \#117016.}
\end{figure}

Similar results for discharge \ \#117016 are shown in Fig.
\ref{4.7}-\ref{4.9}.

\begin{figure}[h]
  \resizebox{1.0\columnwidth}{!}{\includegraphics{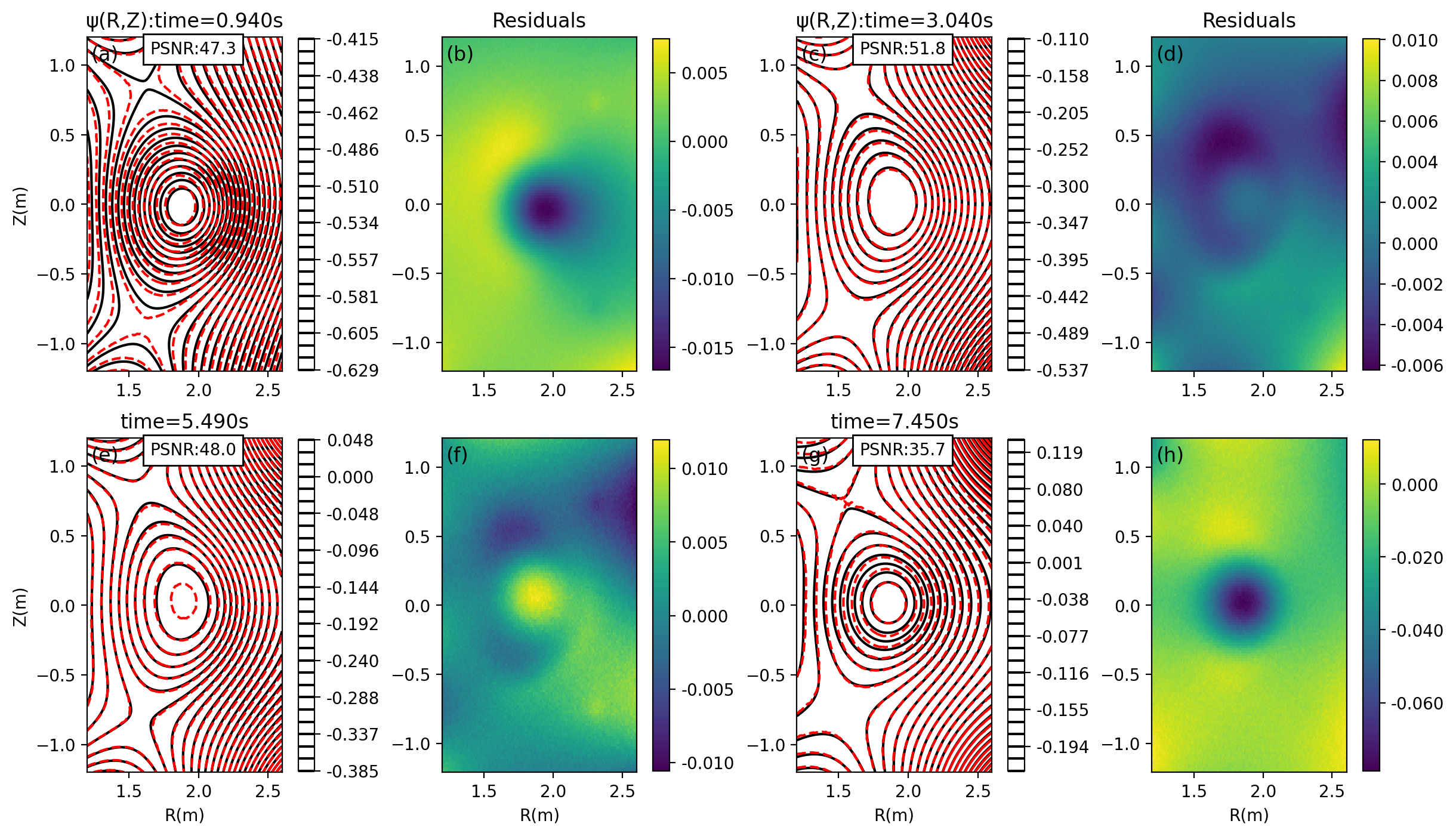}}
  \caption{ Same as figure \ref{4.5}, except that the discharge is \#117016.}
\end{figure}

\begin{figure}[h]
  \resizebox{1.0\columnwidth}{!}{\includegraphics{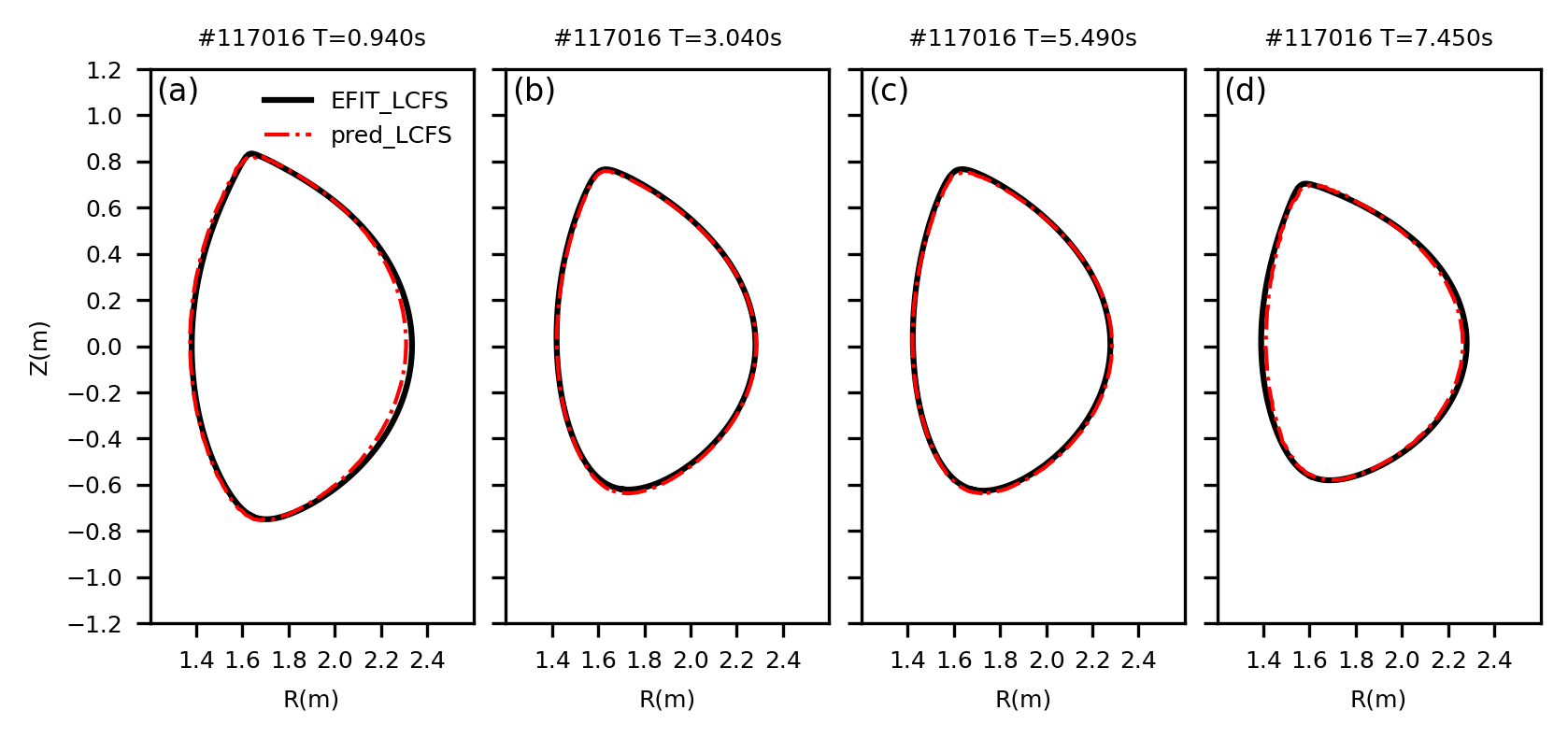}}
  \caption{ \label{4.9}Same as figure \ref{4.6}, except that the discharge is
  \#117016.}
\end{figure}

\begin{figure}[h]
  \resizebox{0.9\columnwidth}{!}{\includegraphics{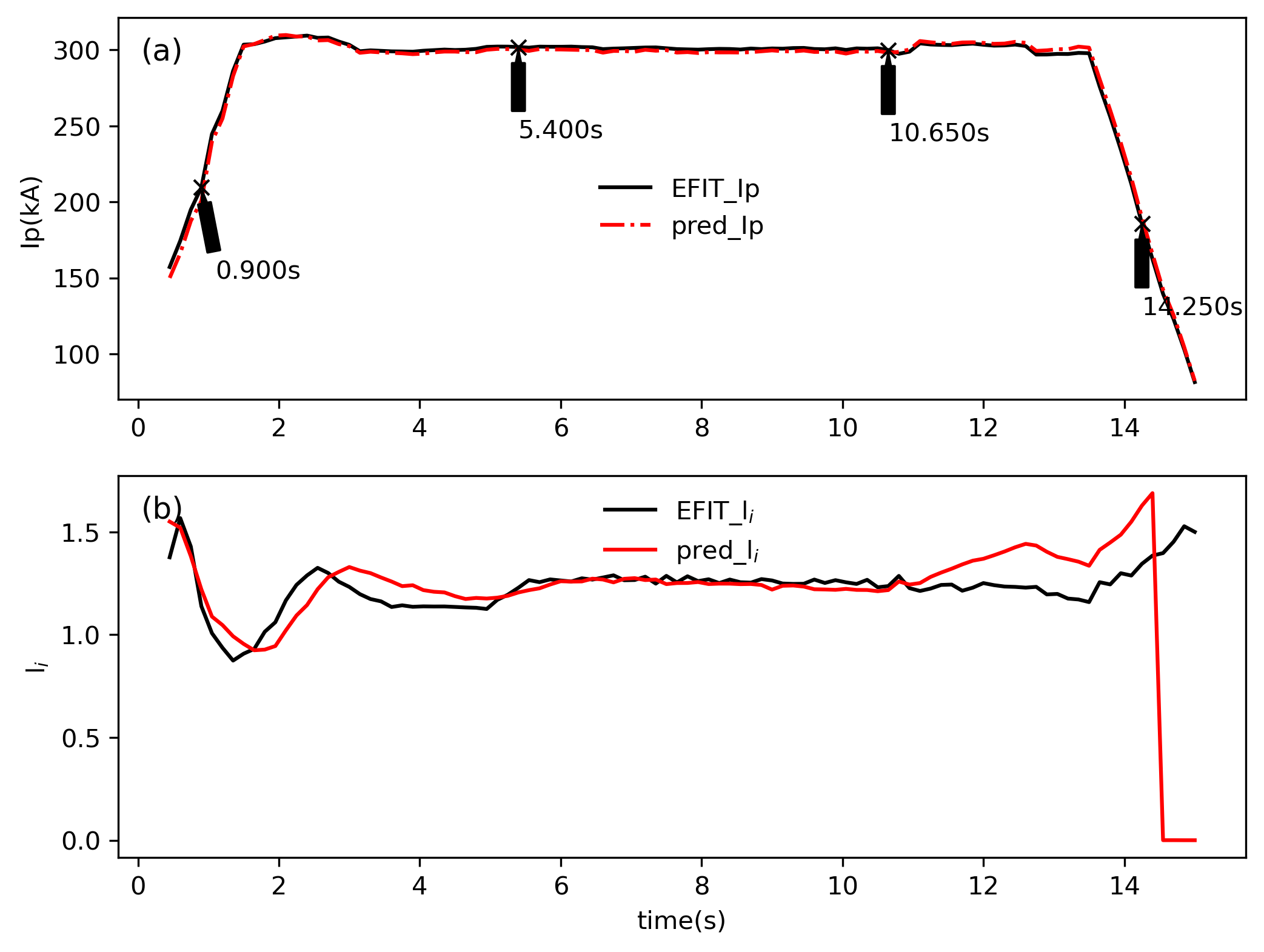}}
  \caption{ \label{4.10}Same as figure \ref{4.4}, except that the discharge is
  \#113019.}
\end{figure}

Similar results for discharge \ \#113019 are shown in Fig.
\ref{4.10}-\ref{4.12}.

\begin{figure}[h]
  \resizebox{1.0\columnwidth}{!}{\includegraphics{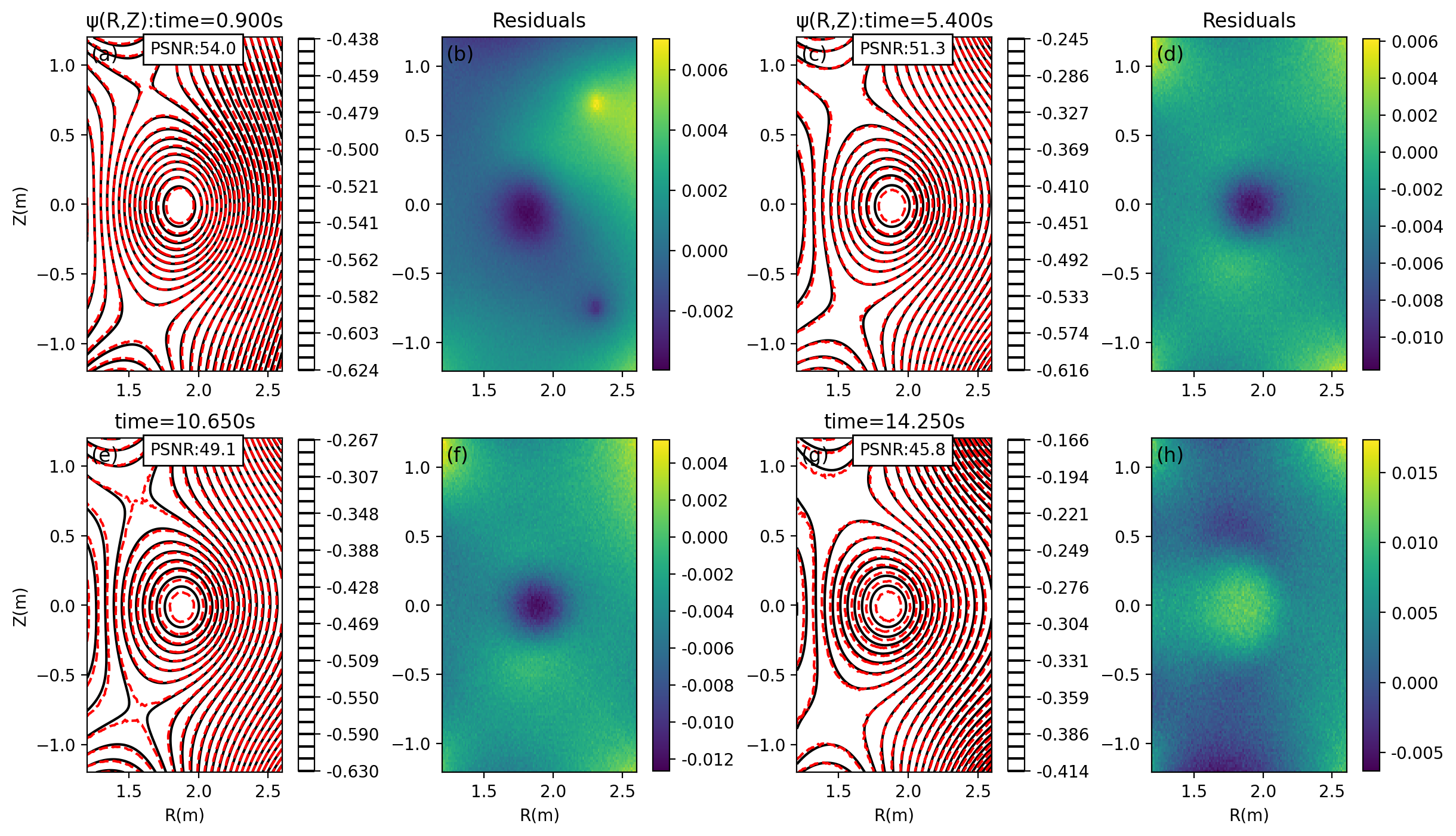}}
  \caption{ Same as figure \ref{4.5}, except that the discharge is \#113019.}
\end{figure}

\begin{figure}[h]
  \resizebox{1.0\columnwidth}{!}{\includegraphics{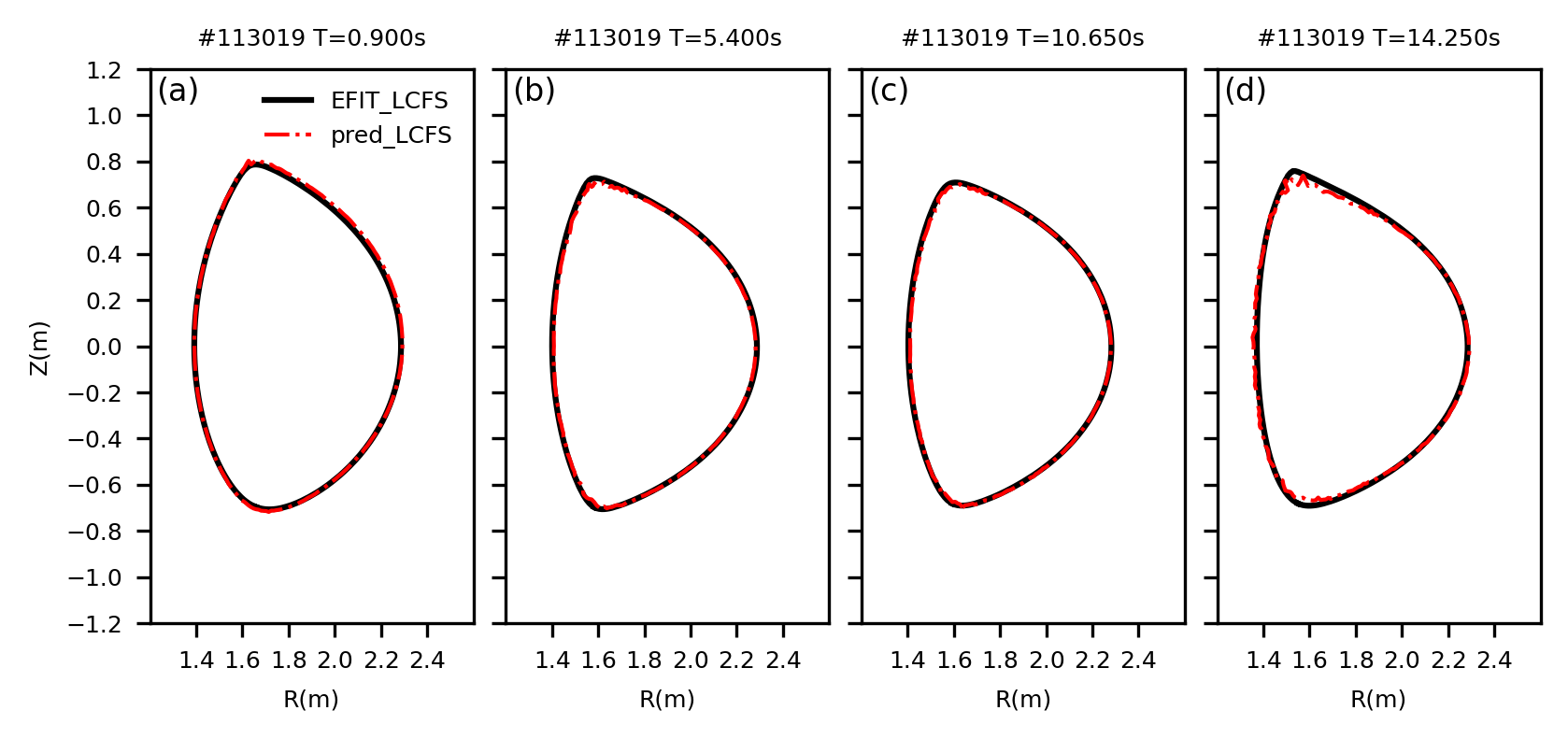}}
  \caption{ \label{4.12}Same as figure \ref{4.6}, except that the discharge is
  \#113019.}
\end{figure}

\section{Neural network prediction of $W_{\tmop{mhd}}$, $\beta_N$, and
$q_{95}$}\label{23-5-18-p4}

Besides the $l_i$ discussed above, there are some other global parameters that
can be constructed from the magnetic measurements, namely the plasma stored
energy $W_{\tmop{mhd}}$, normalized plasma beta $\beta_N$, and edge safety
factor $q_{95}$. These parameters depend on information beyond the poloidal
magnetic flux, namely the toroidal magnetic field and plasma pressure.
Therefore they can not be fully determined by using only the poloidal magnetic
flux predicted from the above network. \ Following Ref. {\cite{DIII-D_2022}},
we construct a new NN for predicting these parameters (called NN2 in the
following; the previous one will be called NN1), where the network has only 3
output values, namely $W_{\tmop{mhd}}$, $\beta_N$, $q_{95}$. The input to NN2
includes a new signal, the current in the toroidal field (TF) coils, which
determines the toroidal field. (In the NN1, this signal is not included
because it has negligible effect on the prediction of the poloidal magnetic
flux.) The NN2 has only one hidden layer consisting of 16 units, and uses the
sigmoid as the activation function for both the hidden and output layers. The
input and output signals of NN2 are normalized by using the same min-max
scalar as used for NN1.

\begin{figure}[h]
  \resizebox{0.9\columnwidth}{!}{\includegraphics{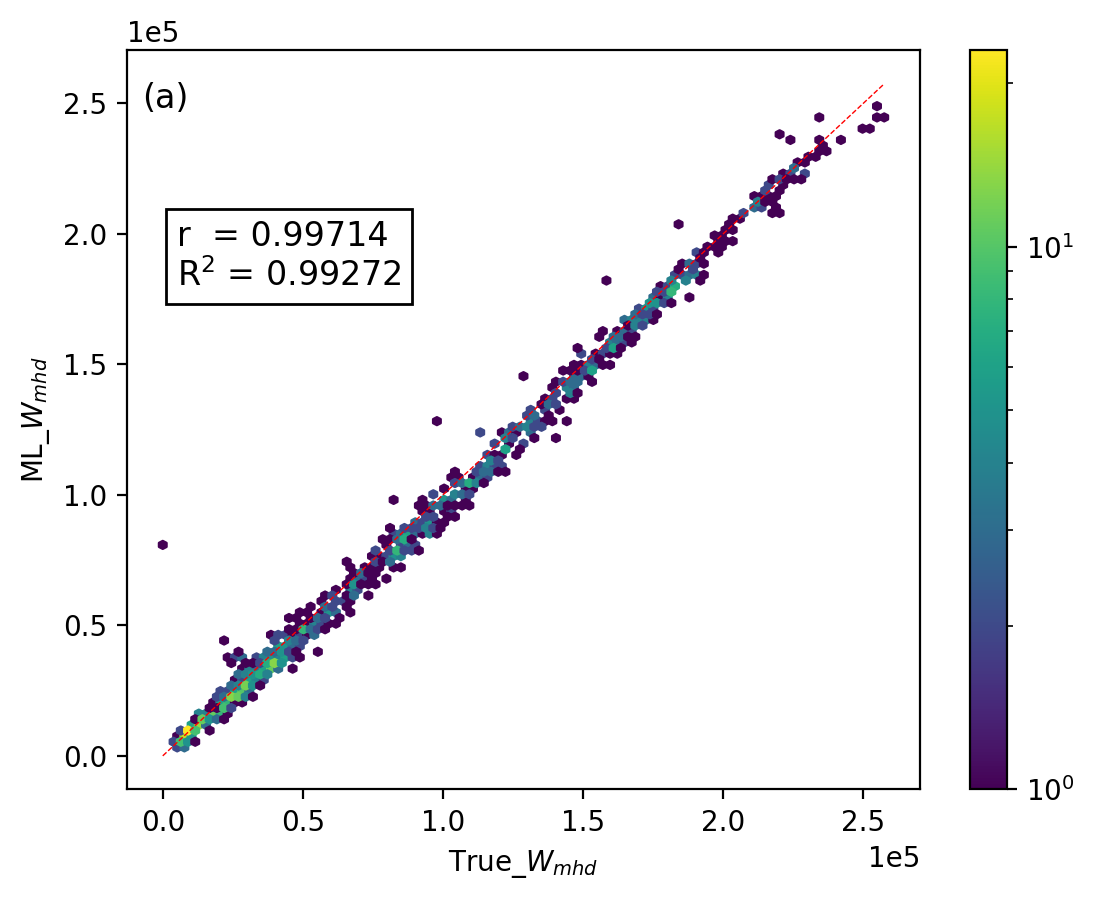}}
  
  \resizebox{0.9\columnwidth}{!}{\includegraphics{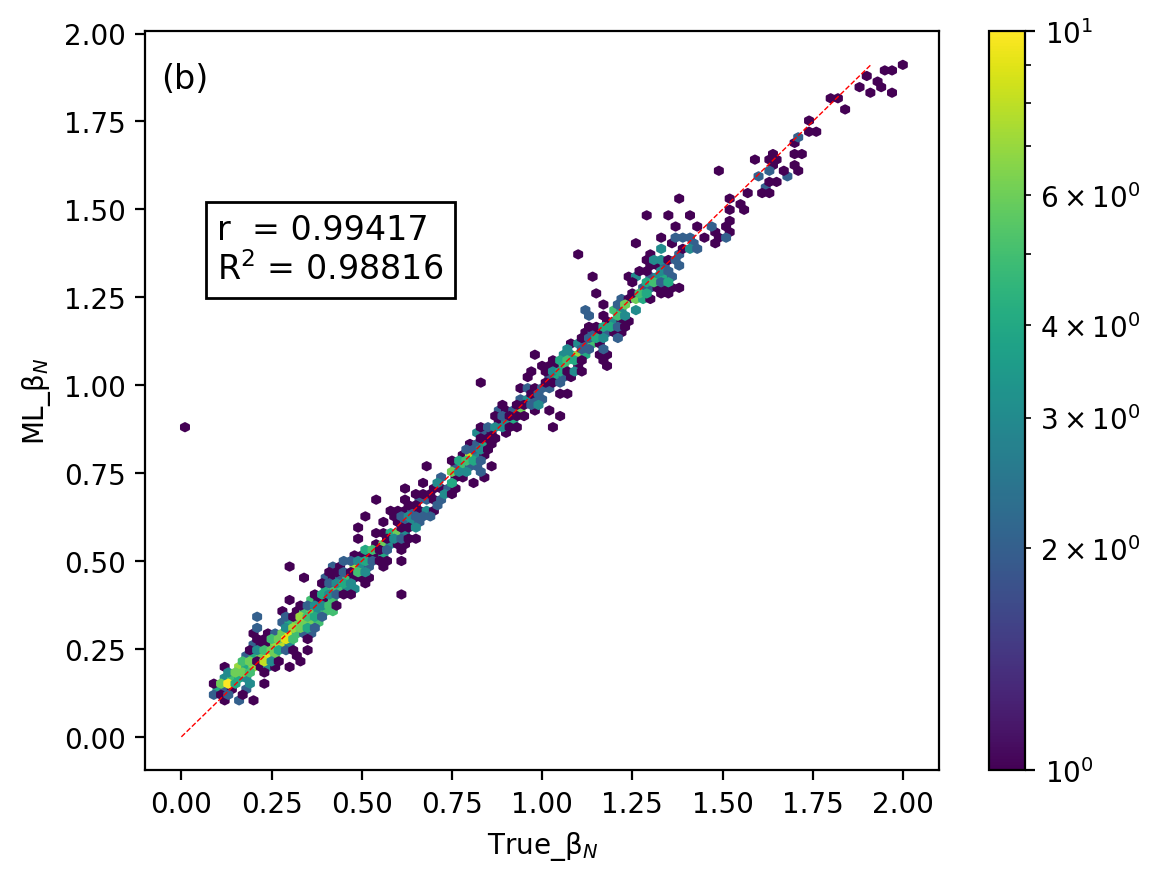}}
  
  \resizebox{0.9\columnwidth}{!}{\includegraphics{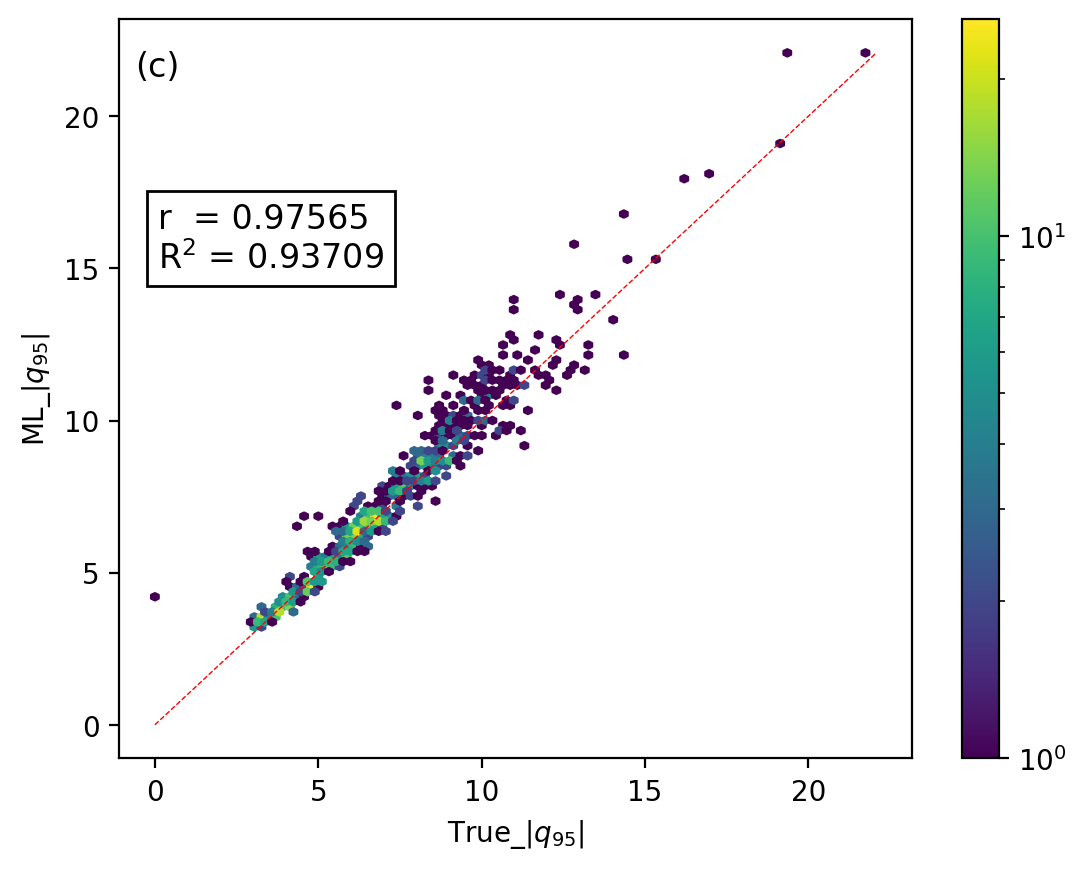}}
  
  \
  \caption{\label{23-5-18-p1}NN2 predictions of $W_{\tmop{mhd}}$, $\beta_N$,
  and $q_{95}$ against the EFIT values for the testing set.}
\end{figure}

The training data consist of about $1 / 4$ randomly selected part of the data
used for NN1. We found that using larger dataset makes this small network
prone to overfitting. The testing set consists of 1000 time slices. Figure
\ref{23-5-18-p1} plots the NN2 predictions against the EFIT values for the
testing set. The results indicate that the NN2 predictions are in reasonable
agreement with the EFIT values for all the 3 parameters. The NN2 predictions
of $q_{95}$ are a little worse than those of the other two parameters, judging
from the values of $r$ and $R^2$.

\begin{figure}[h]
  \resizebox{0.7\columnwidth}{!}{\includegraphics{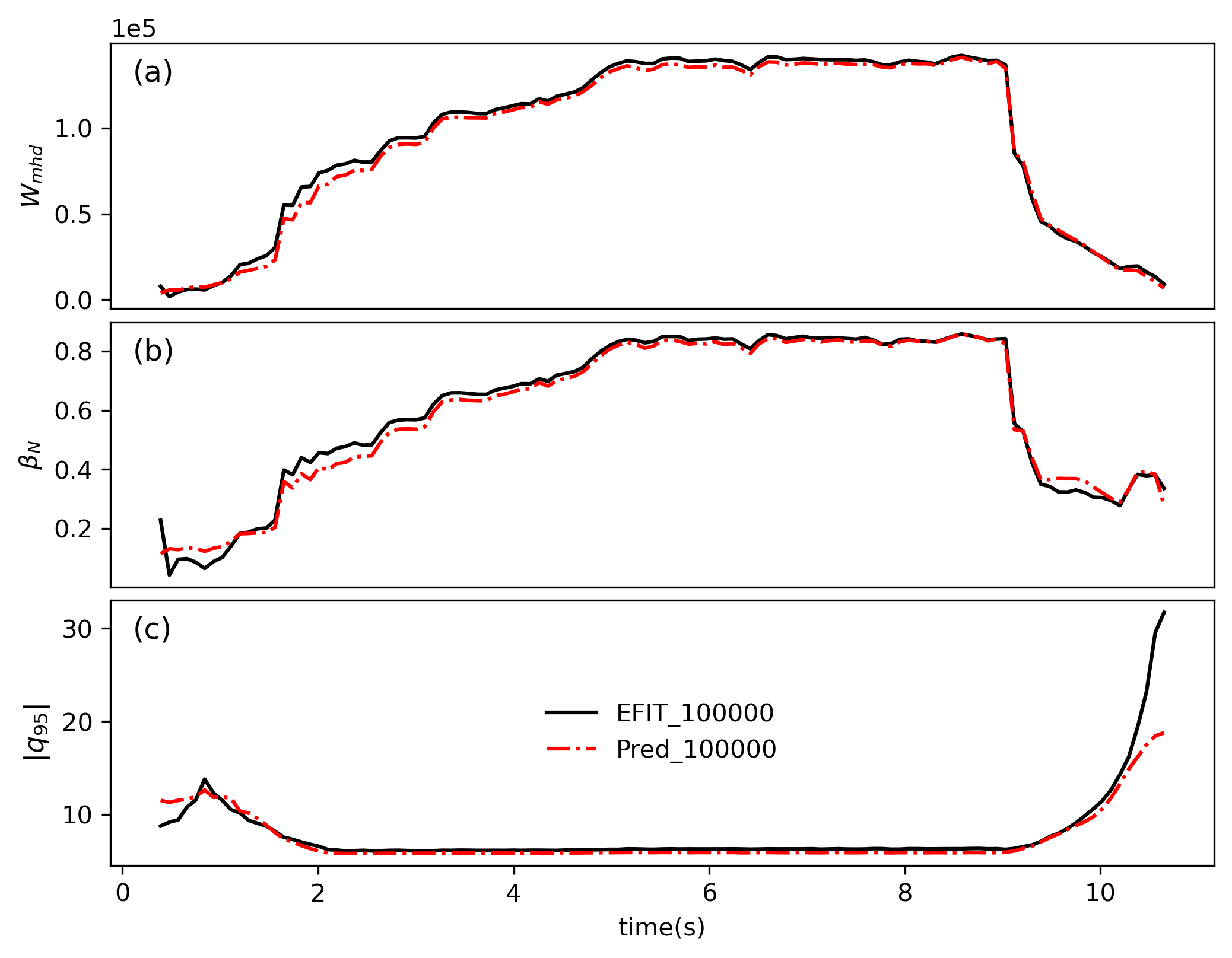}}
  
  \
  
  \
  \caption{\label{23-5-18-p2}Time evolution of plasma stored energy
  $W_{\tmop{mhd}} (J)$, normalized beta $\beta_N$, and edge safety factor
  $q_{95}$ in EAST discharge \#100000. Solid lines are EFIT values and
  dash-dot lines are network predictions.}
\end{figure}

To evaluate the accuracy of NN2 prediction for a full discharge, we
arbitrarily chosen a discharge and compare the time evolution of
$W_{\tmop{mhd}}$, $\beta_N$, and $q_{95}$ between the NN2 predictions and EFIT
values. The results are shown in Fig. \ref{23-5-18-p2}, which shows good
agreement between the network predictions and EFIT values.

\

\section{Summary and discussion}\label{5_summary}

In this work, we train a multiply-layer neural network on the magnetic
measurements (input) and EFIT poloidal magnetic flux (output) on EAST tokamak.
The prediction capability of the network is examined by comparing the
reconstructed magnetic surfaces, last closed flux surfaces, plasma current,
and normalized internal inductance with those of EFIT. The neural network
shows good agreement with EFIT for the data unseen in the training process.

In constructing the neural network, we use automatic optimization in searching
for the best hyperparameters of the model. The hyperparameters found this way
turn out to be better than our previously manually set hyperparameters in
terms of the model accuracy.

Based on the model's good prediction capability and efficiency in terms of
computational time (about 0.5ms per equilibrium on a desktop computer, using
an 11th Gen Intel(R) Core(TM) i5-11500@2.70GHz CPU with a single thread), it
looks promising to apply the neural network to real-time magnetic
configuration control. The above computational time does not include the time
used for tracing boundary/internal magnetic surfaces, and other related
calculations to obtain $l_i$. These computations (not optimized in this work)
seem too inefficient to be used in real-time control. The purpose of computing
$l_i$ for the NN1 model is to evaluate the accuracy of the predicted $\Psi$.
To predict these volume-integrated parameters, one usually uses an additional
small network, as we did in Sec. \ref{23-5-18-p4}, which is efficient enough
for real-time control because the network size is usually very small.

This work is limited to magnetic measurements. We plan to add more diagnostics
relating to the inner safety factor profiles and pressure profiles into the
model, in order to construct more realistic equilibria. This will rely on the
kinetic EFIT output. We are accumulating these kind of training data.

\section{Data availability}

The data that supports the findings of this study are available from the
corresponding author upon reasonable request.

\section{Acknowledgments}

The authors thank Ting Lan, Tonghui Shi, Chengguang Wan, Zhengping Luo, Yao
Huang, Guoqiang Li, and Jingping Qian for useful discussions. This work was
supported by Comprehensive Research Facility for Fusion Technology Program of
China under Contract No. 2018-000052-73-01-001228, by Users with Excellence
Program of Hefei Science Center CAS under Grant No. 2021HSC-UE017, and by the
National Natural Science Foundation of China under Grant No. 11575251.

\subsection{Conflict of interest}

The authors have no conflicts to disclose.

\

\appendix\section{Pearson correlation coefficient $r$}\label{r}

The Pearson correlation coefficient $r$ is a statistical measure used to
assess the strength and direction of a linear relationship between the
predicted and true values of the data. It ranges from -1 to 1, where 1
indicates a perfect positive correlation, 0 indicates no correlation, and -1
indicates a perfect negative correlation. The formula for $r$ is
\begin{equation}
  \label{A.1} r = \frac{\sum^n_{i = 1} (y _i - \bar{y}  ) (\hat{y}_i -
  \overline{\hat{y} } )}{\sqrt{\sum^n_{i = 1} (y _i - \bar{y}  )^2}
  \sqrt{\sum^n_{i = 1} (\hat{y}_i - \overline{\hat{y} })^2}},
\end{equation}
where $n$ is the number of data in the testing set, $y_i$ is the value given
by EFIT, $\hat{y}_i $the prediction by the NN, $\bar{y} = \frac{1}{n}
\sum^n_{i = 1} y _i $ is the mean value of the values given by EFIT, and
$\overline{\hat{y} } = \frac{1}{n} \sum_{i = 1}^n \hat{y}_i $is the mean value
predicted by the NN.

\section{\label{r2}Coefficient of determination $R^2$}

Another relevant metric used to assess how well a model fits the data is the
coefficient of determination $R^2$, which is defined by
\begin{equation}
  R^2 = 1 - \frac{\sum_{i = 1}^n  (y _i - \hat{y}_i )^2}{\sum_{i = 1}^n  (y _i
  - \bar{y} )^2},
\end{equation}
where $n, y_i$, $\hat{y}_i $and $\bar{y}$ \ mean the same as in section
\ref{A.1}. The value of $R^2$ ranges from arbitrary negative values to 1,
where 1 represents a perfect fit between the model predictions and the actual
data points. A higher value of $R^2$ suggests that the model is a better fit
for the data. The coefficient of determination $R^2$ is usually not equal to
the squared Pearson correlation coefficient except in some specific cases.

\section{\label{PSNR}Peak signal-to-noise ratio (PSNR)}

The PSNR is a metric that measures the quality of an image by comparing the
original image to a reconstructed version. A higher PSNR value indicates a
higher quality reconstruction. It is defined by
\begin{eqnarray}
  \tmop{PSNR} & = & 10 \times \log_{10} \left( \frac{\max (y_i)^2}{\tmop{MSE}}
  \right) \nonumber\\
  & = & 10 \times \log_{10} \left( \frac{\max (y_i)^2}{\frac{1}{M} \sum^M_{i
  = 1} (y_i - \hat{y}_i)^2} \right) 
\end{eqnarray}
where $\max (y_i)$ is the maximum value of $\Psi$ given by EFIT in $(R, Z)$
plane, and MSE is the mean squared error between the EFIT and NN.

\

\section{\label{23-3-21-a2}Normalized internal inductance}

The normalized internal inductance $l_i$ is defined by
\begin{equation}
  \label{li} l_i = \frac{\langle B^2_{\theta} \rangle_P}{\langle B^2_{\theta}
  \rangle_S},
\end{equation}
where $P$ is the integration over the plasma volume, $\langle B^2_{\theta}
\rangle_S$ is the surface average of poloidal field over the plasma boundary.
$l_i$ reflects the peakness of the plasma current density profile: a small
value of $l_i$ corresponds to a broad current profile.

For circular cross section with minor radius $a$ and assuming $B_{\theta}$ is
independent of the poloidal angle, then, Ampere's law gives $B_{\theta} (a) =
\mu_0 I / (2 \pi a)$. Then $\langle B^2_{\theta} \rangle_S$ is approximated as
\begin{equation}
  \langle B^2_{\theta} \rangle_S \approx B^2_0 (a) = \frac{{\mu^2_0}_{} I^2}{4
  \pi^2 a^2},
\end{equation}
Using this and noting $V \approx \pi a^2 2 \pi R_0$, where $R_0$ is the major
radius of the device, Eq. (\ref{li}) is written as
\begin{equation}
  \label{23-3-24-p5} l_i = \frac{4 \pi^2 a^2}{{\mu^2_0}_{} I^2} \langle
  B^2_{\theta} \rangle_P = \frac{4 \pi^2 a^2 R_0}{{\mu^2_0}_{} I^2 R_0}
  \langle B^2_{\theta} \rangle_P = \frac{2 V}{{\mu^2_0}_{} I^2 R_0} \langle
  B^2_{\theta} \rangle_P .
\end{equation}
Eq. (\ref{23-3-24-p5}) is used in this work to calculate $l_i$.

\end{document}